\pdfoutput=1
\RequirePackage{ifpdf}
\ifpdf 
\documentclass[pdftex]{sigma}
\else
\documentclass{sigma}
\fi

\numberwithin{equation}{section}

\newtheorem{Theorem}{Theorem}[section]
\newtheorem*{Theorem*}{Theorem}
\newtheorem{Corollary}[Theorem]{Corollary}
\newtheorem{Lemma}[Theorem]{Lemma}

\theoremstyle{definition}
\newtheorem{Definition}[Theorem]{Definition}

\newtheorem{Remark}[Theorem]{Remark}

\def\vt{\vec{t}}
\def\vtp{\vec{t'}}

\def\psistar{\psi^{\star}}

\def\X{\mathcal{X}}
\def\CM{\mathcal{C\!M}}
\def\N{\mathbb{N}}
\def\C{\mathbb{C}}

\def\ff{\hat f}
\def\gg{\hat g}
\def\fff{\hat \ff}
\def\ggg{\hat \gg}

\begin{document}

\allowdisplaybreaks

\newcommand{\arXivNumber}{2511.11813}

\renewcommand{\PaperNumber}{036}

\FirstPageHeading

\ShortArticleName{Orthogonality, KP Wave Functions, and the Bispectral Involution}

\ArticleName{Orthogonality with Respect to the Hermite Product, KP Wave Functions, and the Bispectral Involution}

\Author{Alex KASMAN~$^{\rm a}$, Robert MILSON~$^{\rm b}$ and Michael GEKHTMAN~$^{\rm c}$}

\AuthorNameForHeading{A.~Kasman, R.~Milson and M.~Gekhtman}

\Address{$^{\rm a)}$~College of Charleston, Charleston SC, USA}
\EmailD{\mail{kasmana@cofc.edu}}
\URLaddressD{\url{http://kasmana.people.cofc.edu/}}

\Address{$^{\rm b)}$~Dalhousie University, Halifax NS, Canada}
\EmailD{\mail{robert.milson@dal.ca}}

\Address{$^{\rm c)}$~University of Notre Dame, South Bend IN, USA}
\EmailD{\mail{mgekhtma@nd.edu}}

\ArticleDates{Received November 18, 2025, in final form March 24, 2026; Published online April 14, 2026}

\Abstract{It is well known that for any wave function $\psi(x,z)$ of the KP hierarchy, there is another wave function called its ``adjoint'' such that the path integral of their product with respect to $z$ around any sufficiently large closed path is zero. For the wave functions in the adelic Grassmannian ${\rm Gr}^{\rm ad}$, the bispectral involution which exchanges the role of~$x$ and $z$ also implies the existence of an ``$x$-adjoint wave function'' $\psi^{\star}(x,z)$ so that the product of the wave function, the $x$-adjoint, and the Hermite weight \smash{${\rm e}^{-x^2/2}$} has no residue. Utilizing this, we show that the sequences of coefficient functions in the power series expansion of any KP wave function in ${\rm Gr}^{\rm ad}$ and its image under the bispectral involution at \smash{$t_2=-\frac{1}{2}$} are always ``almost bi-orthogonal'' with respect to the Hermite product. Whether the sequences have the stronger properties of being (almost) orthogonal can easily be determined in terms of KP flows and the bispectral involution. As a special case, the exceptional Hermite orthogonal polynomials can be recovered in this way. This provides both a~generalization of and an explanation of the fact that the generating functions of the exceptional Hermites are certain special wave functions of the KP hierarchy. In addition, one new surprise is that the same KP wave function which generates the sequences of functions is also a generating function for the norms when evaluated at $t_1=1$ and $t_2=0$. The main results are proved using Calogero--Moser matrices satisfying a rank one condition. The same results also apply in the case of ``spin-generalized'' Calogero--Moser matrices, which produce instances of matrix orthogonality.}


\Classification{33C45; 37K10; 42C05; 15A63}

\section{Introduction}

Consider the classical Hermite polynomials $\{h_n(x)\}$ with the generating function
\[
\Psi(x,z)={\rm e}^{xz-\frac{z^2}{2}}=\sum_{n=0}^{\infty} h_n(x)z^n
\]
and the inner product
\[
\langle f,g\rangle_{\rm H}=\int_{-\infty}^{\infty} f\relax{g}{\rm e}^{-\frac{x^2}{2}}{\rm d}x.\]
One can directly compute
 the value of the Hermite product of two copies of the generating function with
independent variables in the second argument
\begin{equation}\label{eqn:wz}
\langle \Psi(x,w),\Psi(x,z)\rangle_{\rm H}=\int_{-\infty}^{\infty} \Psi(x,w)\relax{\Psi(x,z)}{\rm e}^{-x^2/2}{\rm d}x=\sqrt{2\pi}{\rm e}^{wz}.
\end{equation}
Using the power series expansion of the generating function and
linearity of the product, one sees that the coefficient of $w^mz^n$ in $\langle
\Psi(x,w),\Psi(x,z)\rangle_{\rm H}$ would be $\langle h_m,h_n\rangle_{\rm H}$.
It therefore follows from the fact that the right-hand side of \eqref{eqn:wz}
 depends only on the product $wz$ that the Hermite polynomials satisfy
 $\langle h_m,h_n\rangle_{\rm H}=0$ if $m\not=n$. This is one way to
 determine that the Hermite polynomials are orthogonal with respect to
 that inner product.
(Famously,
 the Hermite polynomials are also eigenfunctions for a second-order
 differential operator and satisfy a certain three-term recurrence
 relation, but those facts will not play a significant role in the present investigation.)

 It is possible to multiply the Hermite generating function by a
 rational function so that the product still generates orthogonal functions.
As will be explained in greater detail in
Sections~\ref{sec:exceptionals} and \ref{sec:xingrad},
for each partition $\lambda$ there exists a~function of the form
 \begin{equation}
 \frac{p_{\lambda}(x,z)}{\tau_{\lambda}(x)}\Psi(x,z)=\sum_{n=0}^{\infty}\hat
 h_n^\lambda(x)z^n,\label{eqn:ratform}
\end{equation}
 where $p_{\lambda}$ and $\tau_{\lambda}$ are both polynomials in
 their arguments and the rational functions $\hat h_n^\lambda(x)$ are also
 orthogonal with respect to the same inner product, or equivalently, the polynomials~${\tilde h_n^\lambda(x)=\tau_{\lambda}(x)\hat h_n^{\lambda}(x)}$ are
 orthogonal with respect to the inner product with the exponential
 Hermite weight modified
 by a factor of $\tau_{\lambda}^{-2}(x)$.
The polynomials $\bigl\{\tilde h_n^\lambda(x)\bigr\}$ do not include
polynomials of every degree, and yet they have many of the other
important properties of the classical Hermite polynomials, including
being eigenfunctions for a second-order differential operator.
 It was
for these reasons that they were named ``exceptional Hermite
polynomials'' and have been the subject of much study (see
\cite{2014dry,XHermite1,XHermite2} and references therein). Since
they are not polynomials but are orthogonal with respect to the
Hermite product, we will refer to the quasi-polynomial sequences $\big\{\hat h_n^\lambda(x)\big\}$
simply as
``exceptional
 Hermites''.

An unexpected connection to soliton theory was found
in \cite{KasMil} where it was shown that generating functions
for the exceptional Hermites
are certain special wave functions of the KP hierarchy from ${\rm Gr}^{\rm ad}$.
One question this raises is whether the orthogonality of the
exceptional Hermites is in any way a consequence of the dynamical
equations of the KP hierarchy. Additionally, since all of the other
wave functions from ${\rm Gr}^{\rm ad}$ also have the form \eqref{eqn:ratform},
one may ask whether the sequences they generate also have any
interesting properties with respect to the Hermite product.

The present paper provides positive answers to both questions.
Beginning with the KP hierarchy written in its integral formulation,
we show that the sequence generated by any wave functions in
${\rm Gr}^{\rm ad}$ and the sequence generated by its bispectral dual are
``almost bi-orthogonal''\footnote{The term
 ``almost biorthogonal'' as used in this paper will be defined in Section~\ref{sec:almostdef}.}
with respect to the Hermite product. The orthogonality
of the exceptional Hermites is recovered as a special case of that
more general result.

One additional new result of the present paper is that the same KP wave
functions which are generating functions for families of exceptional
Hermites also play another role related to their orthogonality.
The wave function is a generating function for the exceptional
Hermites orthogonal with respect to $\langle\cdot,\cdot\rangle_{\rm H}$ if
evaluated at $t_1=x$ and $t_2=-\tfrac{1}{2}$.
We show here that a~normalized version of the same wave function
 evaluated at $t_1=1$ and $t_2=0$ instead
 has a~power series expansion whose coefficients are the norms $\big\langle \hat
 h_n^\lambda,\hat h_n^\lambda\big\rangle_{\rm H}$.

Most prior research on exceptional orthogonal polynomials has utilized the language of
Darboux transformations, but here we will instead be using Calogero--Moser matrices
satisfying the rank one condition $\hbox{rank}([X,Z]-I)=1$.
It is clear from Wilson’s two papers on the Adelic Grassmannian that there is
an equivalence between these two approaches \cite{WilsonBisp,WilsonCM}. Presumably, then, it should be possible to
use the Darboux approach to derive the results below. However, the antiderivative formulas are so
simple to write in terms of Calogero--Moser matrices (see Theorem~\ref{thm:xint}), and so we are choosing to use
them exclusively in this paper and will only mention Darboux transformation briefly in the proof of Theorem~\ref{thm:onlyxh}.

If one replaces the rank one
 condition for Calogero--Moser matrices with a higher rank analogue, the results all generalize
to produce examples of \textit{matrix} functions which are
 orthogonal with respect to a Hermite product, as we show in the final section.

\section{Background}

\subsection{The KP hierarchy in its bilinear integral form}
The function $\psi_0\bigl(\vt,z\bigr)$ depending on $z$ and
$\vt=(t_1,t_2,t_3,\ldots)$, defined as
\smash{$
\psi_0\bigl(\vt,z\bigr)={\rm e}^{\sum t_i z^i}
$}
obviously has the properties
\[
\frac{\partial}{\partial t_1}\psi_0=z\psi_0\qquad\hbox{and}\qquad
\frac{\partial^n}{\partial t_1^n}\psi_0=\frac{\partial}{\partial
 t_n}\psi_0.
\]
That makes it the \textit{simplest} example of a KP wave function.
More generally, a function $\psi\bigl(\vt,z\bigr)$
is called a ``KP wave function'' if it is of the form
\begin{equation}
\psi\bigl(\vt,z\bigr)=\bigl(1+O\bigl(z^{-1}\bigr)\bigr)\psi_0\bigl(\vt,z\bigr)\label{eqn:psiasymptotic}
\end{equation}
and there exists a pseudo-differential operator
$\mathcal{L}=\partial+\sum_{n=1}^{\infty}\alpha_n(\vt\,)\partial^{-n}$
(where $\partial=\partial/\partial t_1$) satisfying
\[
\mathcal{L}\psi=z\psi\qquad\hbox{and}\qquad(\mathcal{L}^n)_+\psi=\frac{\partial}{\partial
t_n}\psi.
\]
The
compatibility of those linear equations are equivalent to a hierarchy of integrable nonlinear PDEs.
Among the equations induced are the KP equation modeling surface water
waves, which is why the collection is called the KP hierarchy \cite{GOST,Sato,SegalWilson}.

It is less common to write the KP
hierarchy in its \textit{integral} formulation, but that alternative
is both more compact and of greater relevance to this note. The
existence of the Lax operator~$\mathcal{L}$ from above is equivalent
to requiring that
there exists another function $\psi^*\bigl(\vt,z\bigr)$ having the same
asymptotic form~\eqref{eqn:psiasymptotic} as~$\psi$
satisfying
\begin{equation}
\oint \psi\bigl(\vt,z\bigr)\psi^*\bigl(\vtp,z\bigr){\rm d}z=0,\label{eqn:KPasintegral}
\end{equation}
where the path of integration is any sufficiently large loop and
 $\vtp=(t_1',t_2',\ldots)$ are a~different set of KP time variables
 independent of $\vt$
 (see \cite{DateJimboKashMiwaI,DateJimboKashMiwaIII,DateJimboKashMiwaIV,DateJimboKashMiwaV,DateJimboKashMiwaVI,DateJimboKashMiwaVII}). Equation \eqref{eqn:KPasintegral} is what we will call the
\textit{bilinear integral formulation of the KP hierarchy}.

The function
$\psi^*\bigl(\vt,z\bigr)$ is generally referred to as \textit{the adjoint wave
 function} of $\psi\bigl(\vt,z\bigr)$. However, in this note it might make
more sense to refer to it more specifically as the ``$z$-adjoint'' because
we will also be interested in another function that is a sort of
``$x$-adjoint''.

\begin{Remark}
 Although a KP wave function by definition depends on the
 infinitely-many time variables $\vec t=(t_1,t_2,t_3,\ldots)$ as well
 as the ``spectral variable'' $z$, in most of this paper we will be
 assuming that $t_n=0$ for $n>2$. When that is the case, we will
 suppress the higher time variables as arguments of the function and
 simply write
 $\psi(t_1,t_2,z)$ as an ``abbreviation'' for~${\psi(t_1,t_2, 0,0,\ldots,z)}$.
\end{Remark}

\subsection{Exceptional Hermites}\label{sec:exceptionals}

Let $\lambda=(\lambda_1,\lambda_2,\ldots,\lambda_N)$
be a partition\footnote{In other words, $\lambda_n\in\mathbb{N}\cup\{0\}$, $\lambda_{n+1}\leq
\lambda_n$, and $|\lambda|=\lambda_1+\lambda_2+\cdots+\lambda_N=N$.} of the number $N$ which has length $N$.
One can create exceptional Hermite quasi-polynomials
associated to the choice of such a partition
in
the following way \cite{XHermite1,XHermite2}. Let $K_{\lambda}$ be the ordinary differential
operator in $x$ whose action on an arbitrary function $f(x)$ is
\[
K_{\lambda}(f)=\frac{\hbox{Wr}(h_{N+\lambda_1-1},h_{N+\lambda_2-2},\ldots,h_{N+\lambda_N-N},f)}{\hbox{Wr}(h_{N+\lambda_1-1},h_{N+\lambda_2-2},\ldots,h_{N+\lambda_N-N})}.
\]
(Note that $K_{\lambda}$ is the unique monic differential operator of
order $N$ having the classical Hermite polynomials
$h_{N+\lambda_n-n}(x)$ (for $1\leq n\leq N$) in its kernel.)
Then define
$
\hat h_n^{\lambda}(x)=K_{\lambda}(h_n(x))$.
These are quasi-polynomials (i.e., a collection of rational functions
sharing a common denominator) which are orthogonal\footnote{Their
 orthogonality in the special case of even partitions was established
 in \cite{2014dry} and the general case was shown in \cite{HHV}.
The reader
 might be worried that the product will not make sense if the
 denominator has real roots since it involves integration over the
 real axis. For this reason, this is why many authors have
 considered only the restriction to even partitions for which those
 problems are avoided. However, as has
 been noted before \cite{HHV} and
 we will demonstrate later, this is not a serious obstacle. In
 particular, since it will be shown that the integrand has no residue
 at any of its singularities, one may always consider the integral from $-\infty$ to $\infty$
 to follow any path which avoids those roots, the integral is
 then well-defined, and then the quasi-polynomials are
 orthogonal as claimed.} with respect to the Hermite product
$\langle \cdot,\cdot\rangle_{\rm H}$. Equivalently, their numerators are
polynomials which are orthogonal with respect to an inner product
with the Hermite weight function divided by the square of their
denominators. Moreover, despite the fact that the
numerators do not include polynomials of every degree since $N$
degrees are missing, it is still
possible to write any polynomial as a (possibly infinite) linear
combination of them.

\subsection{Exceptional Hermite generating functions are KP
 wave functions}\label{sec:xingrad}

Despite the similarity of the trivial KP wave function $\psi_0\bigl(\vt,z\bigr)$ and the
generating function~$\Psi(x,z)$ of the classical Hermites,
 it was a surprise that the generating function for each family of
exceptional Hermite quasi-polynomials is a wave function of the KP
hierarchy \cite{KasMil}. In particular, let
\[
\kappa_n(\vt)=\frac{\partial^{N+\lambda_n-n}}{\partial
 z^{N+\lambda_n-n}}\psi_0\bigl(\vt,z\bigr)\bigg|_{z=0}
\]
and define
\smash{$
\psi_{\lambda}\bigl(\vt,z\bigr)=\frac{\hbox{Wr}(\kappa_1,\ldots,\kappa_N,\psi_0)}{z^N\hbox{Wr}(\kappa_1,\ldots,\kappa_N)}$},
where the derivatives in the Wronskian are taken with~respect to~$t_1$.
This is not only a KP wave function in the sense discussed above, it
is also \textit{bispectral} (being a joint eigenfunction for a ring of
differential operators in~$x$ having eigenvalues depending on~$z$ and
a ring of differential operators in~$z$ having eigenvalues depending
on $x$). And, setting $t_1=x$, $t_2=-\tfrac{1}{2}$ and $t_n=0$ for $n\geq3$, it
is also a generating function for the corresponding
exceptional Hermites in the sense that
\[
z^N\psi_{\lambda}\bigl(x,-\tfrac{1}{2},z\bigr)=z^N\psi_{\lambda}\bigl(x,-\tfrac{1}{2},0,\ldots,z\bigr)=\sum \hat h^{\lambda}_n(x)z^n.
\]
 It will be demonstrated below that
 \begin{equation}
 \label{eqn:orthogasintegral}
\int_{-\infty}^{\infty}\psi_{\lambda}\bigl(x,-\tfrac{1}{2},w\bigr)
 \psi_{\lambda}\bigl(x,-\tfrac{1}{2},z\bigr){\rm e}^{-x^2/2}{\rm d}x
 =f(wz)
 \end{equation}
 is a function whose dependence on $w$ and $z$ is of the special form
 which indicates the orthogonality of the sequence of coefficients
 (cf.~\cite{twofacts}). In
 other words, we will re-derive the orthogonality of the exceptional
 Hermites by directly evaluating the Hermite product of those generating
 functions.

 \subsection{Comparing (\ref{eqn:KPasintegral}) and
 (\ref{eqn:orthogasintegral})}

For any partition $\lambda$, $\psi _{\lambda}$ is
both a KP wave function and a generating function for exceptional
Hermites. That it is a KP wave function can be expressed in integral
form through \eqref{eqn:KPasintegral} and the orthogonality of the
exceptional Hermites can be expressed through an integral of the form~\eqref{eqn:orthogasintegral}.

Those integral equations have some things in common. For example, each integrates
a product involving two KP wave functions. Also, although each of the functions in
the product depends on the same variable of integration, other
parameters in the two functions are independent.

They are
also different in some ways. For example, one is integrated with respect to the
spectral parameter $z$ and the other with respect to the spatial
parameter $x=t_1$. Also, the integral in \eqref{eqn:orthogasintegral}
involves not just those wave functions but also the Hermite weight ${\rm e}^{-x^2/2}$.

However, as we will see below, the two differences described above
both vanish when we use the bispectral involution \cite{WilsonBisp} to
exchange the variables $x$ and $z$, since doing so with the second
time-variable taking a non-zero value \textit{fortuitously}
inserts a factor in the form ${\rm e}^{\alpha x^2}$.

\subsection{Calogero--Moser matrices and KP wave functions}

With $[A,B]=AB-BA$ denoting the usual commutator,
let $\CM_N$ denote the set
\[
\CM_N=\big\{\bigl(X,Z,\vec a,\vec b\,\bigr)\mid [X,Z]-I=\vec b\vec a^{\top}\big\}
\]
of $4$-tuples made up of two $N\times N$ matrices and two $N$-vectors which satisfy
the ``rank one condition''
\begin{equation}
 [X,Z]-I=\vec b\vec a^{\top}.\label{eqn:CM}
\end{equation}
 We call $\bigl(X,Z,\vec a,\vec b\,\bigr)\in \CM_N$ ``Calogero--Moser matrices''
 due to their connection to an integrable particle
system studied in the 1970s: for matrices satisfying \eqref{eqn:CM},
the eigenvalues of the time-dependent matrix $X+ntZ^{n-1}$ move according
to the $n$-th Hamiltonian of the Calogero--Moser particle system (see \cite{KKZ,WilsonCM}).

For any $\X=\bigl(X,Z,\vec a,\vec b\,\bigr)\in \CM_N$,
George Wilson \cite{WilsonCM} showed that the function
\begin{equation}
\psi_{\X}\bigl(\vt,z\bigr)=\left(1+\vec a^{\top}\left(\sum_{n=1}^{\infty}
 nt_nZ^{n-1}-X\right)^{-1}(zI-Z)^{-1}\vec b\right)\psi_0\bigl(\vt,z\bigr)
 \label{eqn:psiXZ}
 \end{equation}
is a KP wave function.
The collection of all functions that can be formed in this way (for all~$N\in \N$) are precisely the KP wave
 functions that make up the ``adelic Grassmannian'' ${\rm Gr}^{\rm ad}$.

In this note, we will mostly be interested in the case that all of the
time variables $t_n=0$ for~$n\geq 3$.
When that is understood, we will write the wave function as $\psi_{\X}(x,y,z)$ to
indicate its dependence on the three remaining parameters $x=t_1$, $y=t_2$ and $z$.

Wilson's
\textit{bispectral involution} on ${\rm Gr}^{\rm ad}$ \cite{WilsonBisp}
takes a particularly elegant form when stated in terms of Calogero--Moser matrices.
Consider the involution $\X=\bigl(X,Z,\vec a,\vec b\,\bigr)\mapsto\X^{\flat}=\smash{\bigl(Z^{\top},X^{\top},\vec b,\vec a\bigr)}$
on $\CM_N$.
When only the first time variable $x=t_1$ and the
 spectral parameter are $z$ non-zero,
 this map simply induces an exchange of the two variables
 \cite{WilsonCM} and we have%
\begin{gather}
 \psi_{\X^{\flat}}(x,0,z)=\psi_{\X}(z,0,x).\label{eqn:bi0}
\end{gather}
Note that the formula above requires $t_n=0$ for
$n\geq 2$. If any of the ``higher times'' is non-zero, the
relationship between the wave functions associated to $\X$ and
$\X^\flat$ becomes a bit more complicated. (See Lemma~\ref{lem:bispy}.)

 \begin{Definition}
For each partition $\lambda$, let $\X_{\lambda}$ denote the
Calogero--Moser matrices so that $\psi_{\X_{\lambda}}\bigl(x,-\tfrac{1}{2},z\bigr)=\psi_{\lambda}\bigl(x,-\tfrac{1}{2},z\bigr)$ is the
 generating function for the exceptional Hermites $\big\{\hat h_n^{\lambda}(x)\big\}$
The particular formula for the Calogero--Moser matrices associated to a partition~$\lambda$ is not needed in this note,
but can be found in~\cite{PThesis, PalusoKasman} (see also the appendix of~\cite{WilsonCM}).
 \end{Definition}

 \begin{Remark}
 The map from Calogero--Moser matrices to KP wave functions is not
 one-to-one. There are many different choices of $\vec a $ and
 $\vec b$ for any given pair $(X,Z)$ for which ${\operatorname{rank}([X,Z]-I)=1}$,
 and that choice does not affect the wave function. Moreover, if~$U$
 is an invertible $N\times N$ matrix then $\X=\bigl(X,Z,\vec a,\vec b\,\bigr)$ and
 $U\X U^{-1}=\bigl(UXU^{-1},UZU^{-1},\bigl(U^{-1}\bigr)^{\top}\vec a,U\vec b\,\bigr)$ yield exactly the same wave function
 $\psi_{\X}=\psi_{U\X U^{-1}}$. So, to have a
 bijection between the union of the sets $\CM_N$ over all natural
 numbers $N$ and ${\rm Gr}^{\rm ad}$ one needs to ignore $\vec a$ and $\vec
 b$ and then quotient out
 by this group action.
 \end{Remark}

\section{Notation and terminology}

\subsection{``Almost'' bi-orthogonality}\label{sec:almostdef}

One can say the
sequences of functions $\{f_n(x)\}$ and $\{g_n(x)\}$ are \textit{bi-orthogonal}
with respect to an inner product (or bilinear form) $\langle
\cdot,\cdot\rangle$ if~$
\langle f_m(x),g_n(x)\rangle = \nu_n \delta_{mn}
$,
where $\{\nu_n\}$ is a sequence of constants and $\delta_{mn}$ is the Kronecker delta function. In the case that the
two sequences coincide~($\forall n(f_n(x)=g_n(x))$), we just say the
sequence is orthogonal.

Let us generalize this concept by saying that the sequences are \textit{almost}
bi-orthogonal (cf.~\cite{BaileyDerev})
if there exists a \textit{finite} set of integers
$B\subset\mathbb{Z}$ and a doubly-indexed sequence of constants
$\nu_{b,n}$ such that
\[
\langle f_m(x),g_n(x)\rangle = \sum_{b\in B}\nu_{b,n} \delta_{m,n+b}.
\]
Note that the product of $f_m$ and $g_n$ is non-zero only if the
difference $m-n$ is in $B$, and in particular the product is zero as
long as $|m-n|$ is sufficiently large. In the case that the sequences
coincide, we will say that the sequence is \textit{almost
 orthogonal}.\footnote{The term ``almost orthogonal'' is sometimes used in
 orthogonal polynomial literature to mean something else. In
 particular, it may imply to some readers that only a finite number
 of off-diagonal entries in the matrix of products are non-zero.
 However, here it will be used more generally to say that non-zero
 entries only appear on a finite number of sub- or super-diagonals.} And, in the case~$B=\{0\}$, one has orthogonality as
described in the previous paragraph.

\subsection{Modifications of given Calogero--Moser matrices}\label{sec:tilde}

Suppose $X$ and $Z$ are two $N\times N$ matrices. Then, merely for
convenience, throughout the remainder the
following ``shorthand'' will be used to denote certain specific
matrices made from them
\begin{gather*}
\tilde X=xI+2yZ-X,\qquad
\tilde Z=zI-Z,
\qquad
\acute X=x'I+2y'Z-X,\\
\acute Z=w I+(1-4yy')Z+2y'X,
\qquad
X^\star=(4yy'-1)Z-2y'X,\qquad Z^\star=X-2yZ.
\end{gather*}
Here
$x$, $y$, $z$, $x'$, $y'$, and $w$ are scalar parameters
and $I=I_N$ denotes the $N\times N$ identity matrix.

Furthermore, if $\X=\bigl(X,Z,\vec a,\vec b\,\bigr)$, then
 \[
\X^{\flat}=\bigl(Z^{\top},X^{\top},\vec b,\vec a\bigr),\qquad
\X^*=\bigl(-X^{\top},Z^{\top},\vec b,\vec a\bigr),\qquad\hbox{and}\qquad
\X^\star=\bigl(X^\star,Z^\star,\vec a,\vec b\,\bigr).
 \]

 \begin{Remark}
 The matrix $\tilde X$ will be familiar to anyone who has read
 Wilson's seminal paper~\cite{WilsonCM}, or even earlier papers on
 the Calogero--Moser particle system. It simply represents the time
 evolution of the matrix $X$ under the second integrable flow
 $t_2$. Similarly, $\X^*$ and $\X^{\flat}$ are recognizable as inducing the
adjoint and bispectral involutions on ${\rm Gr}^{\rm ad}$.
The formula for $X^\star$, however, looks odd and may require some explanation. It is
 probably best to think of both $y$ and $y'$ as being amounts translated
 under the second KP flow, but with the bispectral involution being
 applied between them, which keeps us from being able to simply add
 them.
 \end{Remark}

\subsection{Antiderivatives involving the ``error function''}\label{sec:mu}
The strange looking function
\[
 \mu(z)=\frac{\sqrt{\pi }
 {\rm e}^{-\frac{(x+x')^2}{4
 (y+y')}}
 \text{erfi}\left(\frac{x+x'+2 z (y+y')}{2
 \sqrt{y+y'}}\right)}{2
 \sqrt{y+y'}}
 \]
is important, primarily because of its more recognizable derivative
 \[ \frac{\partial\mu}{\partial z}=\mu'(z)
={\rm e}^{(x+x')z+(y+y')z^2}=\psi_0(x,y,z)\psi_0(x',y',z).
 \]
Similarly, the function
\[
\tilde\mu(x)=-\frac{\sqrt{\pi } {\rm e}^{y (w^2+z^2)-\frac{(w+z)^2}{4 y'}}
 \text{erf}\left(\frac{w+2 x y'+z}{2
 \sqrt{-y'}}\right)}{2 \sqrt{-y'}}
 \]
has the derivative
\[
\frac{\partial\tilde\mu}{\partial x}=\tilde\mu'(x)={\rm e}^{(w+z)x+(w^2+z^2)y+y'x^2}=\psi_0(x,y,w)\psi_0(x,y,z){\rm e}^{y'x^2}.
\]

\section[Integrating a product of wave functions with respect to z]{Integrating a product of wave functions with respect to $\boldsymbol{z}$}
If one randomly picks two KP wave functions, there is no reason to
expect their product to have a simple antiderivative. This is true
even if both of the wave functions are from ${\rm Gr}^{\rm ad}$.

Using the notation from Section~\ref{sec:tilde}, for any wave function
in ${\rm Gr}^{\rm ad}$ we can find another
so that their product has a simple antiderivative
and use it to conclude that all of the residues of that product in $z$
are zero:
\begin{Theorem}\label{thm:zint}
Let $\X=\bigl(X,Z,\vec a,\vec b\,\bigr)\in \CM_N$ and $\X^*=\bigl(-X^{\top},Z^{\top},\vec
b,\vec a\bigr)$.
Then, an antiderivative of $\psi_{\X}(x,y,z)\psi_{\X^*}(x',y',z)$
with respect to $z$ is
 \begin{equation}
F(z)=\mu(z)
\bigl(1-2(y+y')\vec a^{\top}\tilde X^{-1}\acute X^{-1}\vec
b\,\bigr)
+\mu'(z)
\vec a^{\top}\tilde X^{-1}\tilde Z^{-1}
\acute X^{-1}\vec b.
 \label{eqn:ADprody}
 \end{equation}
Consequently, for
any closed path $C\subseteq \mathbb{C}$ which avoids the eigenvalues
of $Z$,
\begin{equation}
 \oint_C \psi_{\X}(x,y,z)\psi_{\X^*}(x',y',-z){\rm d}z=0.
 \label{eqn:zintegralRES}
\end{equation}
 \end{Theorem}

\begin{proof}
Multiply out
the expression $\psi_{\X}(x,y,z)\psi_{\X^*}(x',y',z)$ in terms of matrices to get that it is equal to
\begin{equation}
\mu'(z)\bigl(1+\vec a^{\top}\tilde
 X^{-1}\bigl(\tilde X\tilde Z^{-1}+\tilde Z^{-1}\acute X
 +\tilde Z^{-1}\vec b\vec
 a^{\top}\tilde Z^{-1}\bigr)\acute X^{-1}\vec b\,\bigr).\label{eqn:prody}
\end{equation}

It follows from the rank one condition that $[\tilde Z,\acute X]-\vec
b\vec a^{\top}=I$. Multiplying this on the left and right by $\tilde
Z^{-1}$, one finds $\tilde Z^{-1}\vec b\vec a^{\top}Z^{-1}=-\tilde
Z^{-2}+\acute X\tilde Z^{-1}-\tilde Z^{-1}\acute X$. Making that
substitution in~\eqref{eqn:prody} yields
\[
 \mu'(z)\bigl(
 1+\vec a^{\top} \tilde X^{-1}\bigl(\bigl(\tilde X+\acute X\bigr)\tilde
 Z^{-1}-\tilde Z^{-2}\bigr)\acute X^{-1}\vec b\,\bigr)
\]
as an alternative formula for the integrand
$\psi_{\X}(x,y,z)\psi_{\X^*}(x',y',z)$.

Then, use the facts that $\tilde X+\acute
X=((x+x')+2(y+y')z)I-2(y+y')\tilde Z$ and
$\mu''(z)=((x+x')+2(y+y')z)\mu'(z)$ to further rewrite the integrand
in the form
\[
\mu'(z)\bigl(1-2(y+y')\vec a^{\top}\tilde X^{-1}\acute X^{-1}\vec b\,\bigr)
-
\mu'(z)\vec a^{\top}\tilde X^{-1}\tilde Z^{-2}\acute X^{-1}\vec b
+\mu''(z)\vec a^{\top}\tilde X^{-1}\tilde Z^{-1}\acute X^{-1}\vec b.
\]
The claimed antiderivative formula then follows from noting that the first term here is the derivative of the first term of \eqref{eqn:ADprody},
and the other two terms here are the result of applying the product
rule to the second term of \eqref{eqn:ADprody}.

That the path integral is zero follows from the fact that the antiderivative
for $F(z)$ is single valued and hence all residues of the integrand are zero.\end{proof}

\begin{Remark}
The map $\chi\mapsto \chi^*$ on Calogero--Moser matrices induces the
adjoint map on the corresponding KP wave functions. (See \cite[Lemma 7.7]{WilsonCM}.)
It therefore follows from
\eqref{eqn:KPasintegral} that~\eqref{eqn:zintegralRES} would be true so long as the path is
sufficiently large to contain all of the poles. In other words,
we know that the sum of the residues of
the integrand must be zero merely because it is the product of a KP
wave function and its adjoint.
\end{Remark}
\begin{Remark}
Note that \eqref{eqn:zintegralRES} here is a \textit{stronger} statement
than the bilinear integral form of the KP hierarchy.
It is not only the sum of the residues that is zero, but in fact each
of the
residues in $z$ would be zero for the product of any KP wave function
in ${\rm Gr}^{\rm ad}$ and its adjoint.
The explicit antiderivative formula above provides a new proof of this fact which was first established
by Haine and Iliev \cite{HaineIliev} in the context of discrete KP.
\end{Remark}

\section[Integrating a product of two wave functions and a Hermite weight with respect to x]{Integrating a product of two wave functions\\ and a Hermite weight with respect to $ \boldsymbol{x}$}\label{sec:intx}

George Wilson noted that the ($z$-)adjoint for any wave function in
${\rm Gr}^{\rm ad}$ is also in ${\rm Gr}^{\rm ad}$~\cite{WilsonBisp}.
Then, if you combine Theorem~\ref{thm:zint}
with the bispectral involution~\eqref{eqn:bi0}
one can quickly conclude that for every KP wave function in ${\rm Gr}^{\rm ad}$ there is
another so that their product has an elementary antiderivative and
residue zero at every value of $x$. In other words, although most KP
wave functions only have a corresponding adjoint so that their product
satisfies that integral equation in $z$, the ones in ${\rm Gr}^{\rm ad}$ also have the same
property in $x$ as well.

At first it might
appear that this only works when $t_n=0$
for $n\geq 2$ since that is assumed to be true in \eqref{eqn:bi0}. The main point of this section is to see what sort of
result analogous to Theorem~\ref{thm:zint} we can get for integrals of
wave functions in ${\rm Gr}^{\rm ad}$ with respect to $x$ when $y=t_2$ is not
assumed to be zero.

Before we state the results, it may be instructive to consider why the
bispectral involution \textit{cannot} be as simple if any of the
higher times are non-zero. Recall that each wave function in~${\rm Gr}^{\rm ad}$ is of the form
\[
\psi_{\X}(x,y,z)={\rm e}^{xz + y z^2}R_{\X}(x,y,z),
\]
where $R_{\X}$ is a rational function.
The important point is that when $y=0$, both $\psi_{\X}(x,0,z)$ and~$\psi_{\X}(z,0,x)$
have the same general form of a rational function times ${\rm e}^{xz}$.
Then, it makes sense that (and thanks to Wilson \cite{WilsonBisp,WilsonCM} we
know it is \textit{true} that) for every $\X$ there is an $\X^{\flat}$
so that $\psi_{\X}(z,0,x)=\psi_{\X^{\flat}}(x,0,z)$. On the other
hand, when $y\not=0$ then swapping the variables~$x$ and $z$ yields
something which could not possibly be equal to the wave function of
any point in~${\rm Gr}^{\rm ad}$ because it has a factor of ${\rm e}^{x^2y}$
\[
\psi_{\X}(z,y,x)={\rm e}^{xz+x^2y}R_{\X}(z,y,x)\not=\psi_{\X'}(x,y,z) \qquad\text{for
any}\ \X'\in\CM_N.
\]
It is for precisely this reason that
 the simplest
generalization of the bispectral involution to the case of non-zero $y$ requires the
introduction of a
factor that looks like a Hermite weight function.

\begin{Lemma}\label{lem:bispy}
For $\X=\bigl(X,Z,\vec a,\vec b\,\bigr)\in\CM_N$, let
$\X^{\flat}(y)=\bigl(Z^{\top},X^{\top}-2yZ^{\top},\vec b,\vec a\bigr)$ denote
the Calogero--Moser matrices obtained by following the second time flow
for $y$ units of time and then applying the bispectral involution.
The following identity expresses the
relationship between the bispectral involution on CM matrices and the
exchange of parameters $x$ and $z$ in the wave function if $t_2=y$ is not zero
\smash{$
\psi_{\X}(z,y,x)={\rm e}^{yx^2}\psi_{\X^{\flat}(y)}(x,0,z)$}.
\end{Lemma}
\begin{proof}
Simply swapping the $x$ and $z$ in the usual definition of the wave
function, we get
\[
 \psi_{\X}(z,y,x)={\rm e}^{xz+x^2y}\bigl(1+\vec
 a^{\top}(zI+2yZ-X)^{-1}(xI-Z)^{-1}\vec b\,\bigr).
\]
On the other hand, if we compute the stationary wave function of
$\X^{\flat}(y)$, we get
\[
\psi_{\X^\flat(y)}(x,0,z)={\rm e}^{xz}\bigl(1+\vec b^\top
 \bigl(xI-Z^\top\bigr)^{-1}\bigl(zI-X^\top+2yZ^\top\bigr)^{-1}\vec a\bigr).
\]
Since that is a scalar, it is equal to its transpose, and so we also
have
\[
\psi_{\X^\flat(y)}(x,0,z)={\rm e}^{xz}\bigl(1+\vec a^\top
 (zI-X+2yZ)^{-1}(xI-Z)^{-1}\vec b\,\bigr).
\]
If you multiply that by ${\rm e}^{yx^2}$ it is equal to the
expression we found earlier for $\psi_{\X}(z,y,x)$, which is what the
claim says.
\end{proof}

The proof of this lemma is merely a straightforward computation, but
 the result happens to be quite useful for our purpose of studying what
 happens when one takes the Hermite product of these wave functions.
 We now introduce the concept of the ``$x$-adjoint'' of a wave function
 in~${\rm Gr}^{\rm ad}$.

\begin{Definition}\label{def:star}
Let $\X=\bigl(X,Z,\vec a,\vec b\,\bigr)\in\CM_N$
and \[\X^{\star}=\bigl(X^\star,Z^{\star},\vec a,\vec b\,\bigr)=
\bigl((4yy'-1)Z-2y'X,X-2yZ,\vec a,\vec b\,\bigr).
\]
Define
\[
\psistar_{\X}(x,y,y',w):={\rm e}^{yw^2}\psi_{\X^{\star}}(w,0,x)={\rm e}^{xw+yw^2}\bigl(1+\vec
a^T\bigl(wI-X^\star\bigr)^{-1}\bigl(xI-Z^\star\bigr)^{-1}\vec b\,\bigr).
\]
We will call this function the $x$-adjoint of $\psi_{\X}(x,y,z)$.
\end{Definition}

By combining Theorem~\ref{thm:zint} and Lemma~\ref{lem:bispy} (and
changing the names of the variables as needed), we get that the
product of any wave
function in ${\rm Gr}^{\rm ad}$ with its $x$-adjoint also satisfies integral equations in $x$.

\begin{Theorem}\label{thm:xint}
Using the notation from Sections~{\rm\ref{sec:tilde}}--{\rm\ref{sec:mu}}, for any values of $y$, $y'$, $w$, and $z$ an antiderivative with
respect to $x$ of the product
\smash{$
\psistar_{\X}(x,y,y',w)\psi_{\X}(x,y,z){\rm e}^{y'x^2}
$}
is
\begin{equation}\label{eqn:ADinx}
\tilde F(x)=\tilde\mu(x)\bigl(1-2y'\vec a^{\top}\acute Z^{-1}\tilde Z^{-1}\vec
b\,\bigr)+\tilde\mu'(x)\bigl(\vec a^{\top}\acute Z^{-1}
\tilde X^{-1}\tilde Z^{-1}\vec b\,\bigr).
\end{equation}
Consequently,
\[
\oint_C\psistar_{\X}(x,y,y',w) \psi_{\X}(x,y,z){\rm e}^{y'x^2}{\rm d}x=0
\]
over any closed loop $C\subseteq\C$ which avoids the poles {\rm(}i.e., the
residues are all zero{\rm)}.
\end{Theorem}

\begin{proof}
As indicated, one way to prove this result is by applying the
bispectral involution to the integral form of the bilinear KP
hierarchy. Alternatively, it can be proved directly using linear
algebra as follows:

 If we multiply the integrand by
$(\tilde \mu'(x))^{-1}$ to cancel out all of the exponential
terms, we are left with
\begin{gather*}
 (\tilde \mu'(x))^{-1}\psistar_{\X}(x,y,y',w) \psi_{\X}(x,y,z){\rm e}^{y'
 x^2}\\
 \qquad = \bigl(1+\vec a^{\top}\acute Z^{-1}\tilde
 X^{-1}\vec b\,\bigr)\bigl(1+\vec a^{\top}\tilde X^{-1}\tilde
 Z^{-1}\vec b\,\bigr)\\
\qquad = 1+\vec a^{\top}\bigl(\tilde X^{-1}\tilde Z^{-1}+\acute Z^{-1}\tilde
 X^{-1}\\
 \phantom{\qquad =}{}
 +\acute Z^{-1}\tilde X^{-1}\vec b\vec a^{\top}\tilde X^{-1}\tilde
 Z^{-1}\bigr)\vec b\qquad\hbox{(just expanding the product)}\\
\qquad = 1+\vec a^{\top}\bigl(\tilde X^{-1}\tilde Z^{-1}+\acute Z^{-1}\tilde
 X^{-1}+\acute Z^{-1}\bigl(-\tilde X^{-2}+[\tilde X^{-1},\acute Z]\bigr)\tilde
 Z^{-1}\bigr)\vec b\qquad\hbox{(using \eqref{eqn:CM})}\\
\qquad = 1 +\vec a^{\top}\bigl(\acute Z^{-1}\tilde X^{-1}-\acute Z^{-1} \tilde
 X^{-2}\tilde Z^{-1}+\acute Z^{-1}\tilde X^{-1}\acute Z\tilde
 Z^{-1}\bigr)\vec b\\
 \qquad = 1 +\vec a^{\top}\acute Z^{-1}\bigl(\tilde X^{-1}\bigl(\tilde Z+\acute
 Z\bigr)-\tilde X^{-2}\bigr)\tilde Z^{-1}\vec b\\
\qquad = 1+ \vec a^{\top}\acute Z^{-1}\bigl((w+z+2xy')\tilde X^{-1}-2y' I_n-\tilde
 X^{-2}\bigr)\tilde Z^{-1}\vec b.
\end{gather*}

\noindent So, the integrand $\psistar_{\X}(x,y,y',w) \psi_{\X}(x,y,z){\rm e}^{y'
 x^2}$ equals
\[
\tilde\mu'(x)\bigl(1+ \vec a^{\top}\acute Z^{-1}\bigl((w+z+2xy')\tilde X^{-1}-2y' I_n-\tilde
X^{-2}\bigr)\bigr)\tilde Z^{-1}\vec b\,\bigr).
\]
However, this is precisely what you get when you compute
\[
\frac{\rm d}{{\rm d}x}\bigl(\tilde\mu(x)\bigl(1-2y'\vec a^{\top}\acute Z^{-1}\tilde
 Z^{-1}\vec b\,\bigr)+\tilde\mu'(x)\bigl(\vec a^{\top}\acute Z^{-1}
 \tilde X^{-1}\tilde Z^{-1}\vec b\,\bigr)\bigr)
\]
because
$\vec a$, $\vec b$, $\tilde Z$, and $\acute Z$ are all independent of $x$,
\[
\frac{\rm d}{{\rm d}x}\tilde X^{-1}=-\tilde X^{-2} \qquad \text{and}\qquad
\frac{\rm d}{{\rm d}x}\tilde \mu'(x)=(w+z+2xy')\tilde \mu'(x).
\]

A path integral of the integrand would be equal to the difference in
the values of that single-valued antiderivative at the endpoints, and in particular
the integral around any closed loop would be zero.
\end{proof}

\begin{Remark}
So, \smash{$\psistar_{\X}\psi_{\X} {\rm e}^{y'x^2}$} has no residues at any of its
poles in $x$ in the same way that $\psi_{\X}\psi^*_{\X}$ has no
residues at any of its poles in $z$.
That is why the similar terminology and notation have been chosen for
the $x$-adjoint $\psistar_{\X}$.
\end{Remark}

\section{Orthogonality}\label{sec:scalarorthog}\label{sec:almostbiorthog}

\subsection[Fixing the value of y']{Fixing the value of $\boldsymbol{y'}$}

Using $\tilde F(x)$ from \eqref{eqn:ADinx} as an antiderivative, we
can say that
\begin{equation}
\int_{-\infty}^{\infty}
\psistar_{\X}(x,y,y',w) \psi_{\X}(x,y,z) {\rm e}^{y'x^2}{\rm d}x=\bigl(\lim_{x\to\infty}
\tilde F(x)\bigr)-\bigl(\lim_{x\to-\infty}\tilde F(x)\bigr).
 \label{eqn:defint}
\end{equation}
This will only be finite in the case $y'<0$. In fact, since we will
be considering this integral as an inner product with weight function
${\rm e}^{y'x^2}$, changing the value of $y'$ will be nothing other than a
scaling of the variable $x$. So, for simplicity we might as well give
$y'$ a specific negative value. The results to follow are
easiest to state if we fix $y'$ as follows:
\[
\text{For the remainder of the paper, we will assume $y'=-\frac{1}{2}$.}
\]
\begin{Remark}
 Note that the integrand may have singularities at real values of
 $x$. One way to make sense of the definite integral in
 \eqref{eqn:defint} in that case is to consider it as a path integral along
 a path that asymptotically approaches the $x$-axis at $x=\pm\infty$,
 but avoids any of the singularities of the integrand. The fact that
 all of the residues are zero guarantees that the value of the
 integral does not depend on the particular path chosen (cf.~\cite{HHV}).
\end{Remark}
\subsection[Hermite product of a wave function
 and its x-adjoint]{Hermite product of a wave function
 and its $\boldsymbol{x}$-adjoint}

\begin{Theorem}\label{thm:scalarip} The Hermite product\footnote{This
 Hermite product is an inner product as long as the integrand remains
 real-valued. However, $\CM_N$~includes complex-valued matrices
 and so that might not be the case. Rather than going through the
 exercise of adding complex conjugation where necessary, we will
 simply extend this formula unchanged from the real to the complex
 case.
 The study of orthogonality with respect to such bilinear forms is
 sometimes referred to as ``non-Hermitian
 orthogonality'' \cite{BarYat}.}
 of the wave
 function corresponding to $\X\in\CM_N$ and
 its $x$-adjoint is
\begin{equation} \label{eqn:scalarip}\bigl\langle
\relax{\psistar_{\X}\bigl(x,y,-\tfrac{1}{2},w\bigr)},\psi_{\X}(x,y,z)\bigr\rangle_{\rm H}=
 \sqrt{2\pi } {\rm e}^{y
 (w^2+z^2)+\frac{
 1}{2} (w+z)^2}
 \bigl(1+\vec a^{\top}\acute Z^{-1}\tilde Z^{-1}\vec b\,\bigr).
\end{equation}
\end{Theorem}
\begin{proof}
We can compute this inner product as the difference between two limits
in \eqref{eqn:defint}.
Since $\tilde\mu'(x)={\rm e}^{-x^2+\alpha x+\beta}$ (where $\alpha$ and
$\beta$ depend on $y$, $w$ and $z$), its limit as $x\to\pm\infty$
goes to zero. The rational function $\vec a^{\top}\acute
Z^{-1}\tilde X^{-1}\tilde Z^{-1}\vec b$ also vanishes at $x=\pm\infty$
(and in any case the product of this rational function with ${\rm e}^{-x^2}$
does). This eliminates the second term on the right of~\eqref{eqn:ADinx} from both limits.

To compute the limits of the first term, note that
\[
\bigl(\lim_{x\to\infty}\tilde\mu(x)\bigr)-
\bigl(\lim_{x\to-\infty}\tilde\mu(x)\bigr)=
 \sqrt{{2\pi} } {\rm e}^{y
 (w^2+z^2)+\frac{
 1}{2} (w+z)^2}.
 \]

The matrix product that $\tilde\mu(x)$ is multiplied by is independent of $x$,
and so the limit in~\eqref{eqn:defint} is simply equal to the limit
found above multiplied by that matrix (now written with~${y'=-\frac{1}{2}}$).
\end{proof}

\begin{Remark}\label{rem:yval}
 Note the expression \[y
 \bigl(w^2+z^2\bigr)+\frac{
 1}{2} (w+z)^2=wz+\left(\frac12+y\right)\bigl(w^2+z^2\bigr)
\] in the exponent of equation~\eqref{eqn:scalarip}. If we are
interested in orthogonality, we would want this expression to be a
function of the product $wz$ (cf.~\cite{twofacts}). This would be the case if and only if
$y=-\frac{1}{2}$.
\end{Remark}

\subsection[Two special values of y and the bispectral involution]{Two special values of $\boldsymbol{y}$ and the bispectral involution}

We have already fixed the value $y'=-\frac{1}{2}$. We need
only consider \textit{two} values of the variable $y$ for the
remainder of the paper, each of which has a special property with
respect to the bispectral involution $\X=\bigl(X,Z,\vec a,\vec b\,\bigr)\mapsto
\X^{\flat}=\bigl(Z^{\top},X^{\top},\vec b,\vec a\bigr)$.

One of those values is $y=0$, which has the special property
\eqref{eqn:bi0}.
The other special value is~${y=-\tfrac{1}{2}}$ which has a surprising
significance when combined with the bispectral involution.
If~we consider only the first two time variables of the KP hierarchy,
the usual $\tau$-function associated to $\X\in\CM_N$ is (cf.~\cite{WilsonCM})
$
\tau_{\X}(x,y)=\det(xI+2yZ-X)$.
Note that $\tau_{\X}(x,0)$ is the characteristic polynomial of
 $X$ and $\tau_{\X^{\flat}}(x,0)$ is the characteristic polynomial of~$Z$. The eigenvalues of $X$ and~$Z$ can be chosen independently
giving the impression that these two $\tau$-functions are not closely related.
However, those two functions are \textit{always} equal at this
 special value of the second time variable.

\begin{Lemma}\label{lem:weirdfact}
For any $\X=\bigl(X,Z,\vec a,\vec b\,\bigr)\in\CM_N$,
 $
 \tau_{\X}\bigl(x,-\frac{1}{2}\bigr)=\tau_{\X^\flat}\bigl(x,-\frac{1}{2}\bigr).
 $\end{Lemma}
 \begin{proof}
Combining the formula above for $\tau_{\X}$ with the bispectral
involution, we get
\[\tau_{\X^\flat}(x,y)=\det\bigl(x I+2y{X}^\top-{Z}^\top\bigr).\]
Substituting $y=-\tfrac{1}{2}$ turns them into
$\det(xI-X-Z)$ and $\det\bigl(xI-X^{\top}-Z^{\top}\bigr)$, which are equal since
the determinant is not affected by taking the transpose of a~matrix.
\end{proof}

As noted in Remark~\ref{rem:yval}, another special
thing about this value is that the exponential term in
\eqref{eqn:scalarip} simplifies when $y'=-\tfrac{1}{2}$. Moreover, there is something nice
to say about the $x$-adjoint wave function
$\psistar_{\X}(x,y,y',w)$ at $y=y'=-\tfrac{1}{2}$.

\begin{Lemma}\label{lem:starbisp}
 For any $\X=\bigl(X,Z,\vec a,\vec b\,\bigr)\in\CM_N$ one has
 \[
 \psistar_{\X}\bigl(x,-\tfrac{1}{2},-\tfrac{1}{2},w\bigr)=\psi_{\X^{\flat}}\bigl(x,-\tfrac{1}{2},w\bigr).
 \]
 That is, for these special values of $y$ and $y'$, the $x$-adjoint
 is the same as the wave function of the image under the bispectral involution.
\end{Lemma}
\begin{proof}
By Definition~\ref{def:star}, when $y=y'=-\tfrac{1}{2}$, then
$X^{\star}=(4yy'-1)Z-2y'X=X$ and $Z^{\star}=X-2yZ=X+Z$. So,
\begin{align*}
\psistar_{\X}\bigl(x,-\tfrac{1}{2},-\tfrac{1}{2},w\bigr)&={\rm e}^{-w^2/2}\psi_{\X^{\star}}(w,0,x)\\
&={\rm e}^{-w^2/2}{\rm e}^{xw}\bigl(1+\vec
 a^{\top}(wI-X^{\star})^{-1}\bigl(xI-Z^{\star}\bigr)^{-1}\vec b\,\bigr)\\
&={\rm e}^{xw-w^2/2}\bigl(1+\vec
 a^{\top}(wI-X)^{-1}(xI-X-Z)^{-1}\vec b\,\bigr)\\
&={\rm e}^{xw-w^2/2}\bigl(1+\vec
 b^{\,\top}\bigl(xI-X^{\top}-Z^{\top}\bigr)^{-1}\bigl(wI-X^{\top}\bigr)^{-1}\vec a\bigr)\\
&={\rm e}^{xw+yw^2}\bigl(1+\vec b^{\,\top}\bigl(xI+2yX^{\top}-Z^{\top}\bigr)^{-1}\bigl(wI-X^{\top}\bigr)^{-1}\vec a\bigr)\\
&=
\psi_{\X^{\flat}}\bigl(x,-\tfrac{1}{2},w\bigr).\tag*{\qed}
\end{align*}\renewcommand{\qed}{}
\end{proof}

Combining Lemma~\ref{lem:starbisp} with Theorem~\ref{thm:xint}, we get
the interesting conclusion the following.
\begin{Corollary}
 For every $\X\in\CM_N$, every $w,z\in\mathbb{C}$, and every closed
 path avoiding the poles of the integrand,
 \[
 \oint \psi_{\X^{\flat}}\bigl(x,-\tfrac{1}{2},w\bigr)\psi_{\X}\bigl(x,-\tfrac{1}{2},z\bigr){\rm e}^{-x^2/2}{\rm d}x=0.
 \]
 \end{Corollary}

As a consequence of the results above, we will be able to conclude
that the sequences of functions generated by any wave function in
${\rm Gr}^{\rm ad}$ and its image under the bispectral involution have
interesting properties relative to the Hermite inner product with
weight ${\rm e}^{-x^2/2}$, and the orthogonality of the exceptional Hermites
will be derived as a special case.

\begin{Remark}
Because the effect of Wilson's bispectral involution on the wave
function is often thought of as simply exchanging $x$ and $z$, one may
mistakenly think that Lemma~\ref{lem:starbisp} says that the wave
function and its $x$-adjoint differ by just such an exchange of
variables. However, that is not quite correct because of the non-zero
value of $t_2$.
\end{Remark}

\begin{Remark}
There are different versions of the classical Hermite polynomials
which are essentially equivalent.
Nevertheless,
the choice of the ``probabilist's''
version with weight function~${\rm e}^{-x^2/2}$ and the corresponding value $\smash{t_2=-\tfrac{1}{2}}$ were not chosen
for this paper arbitrarily. In previous investigations, we
used the ``physicist's'' weight function ${\rm e}^{-x^2}$ and the orthogonal functions
arose from setting $\smash{t_2=-\tfrac{1}{4}}$. We began this investigation using
those choices, but soon realized that the statements and proofs of
theorems would be greatly simplified if we changed those conventions.
From the point of view of the orthogonal functions, changing the value
of $t_2$ amounts to nothing more than a
rescaling of the variable $x=t_1$, as discussed in~\cite{KasMil}.
One could apply such a rescaling to any of the results in this paper
to obtain corresponding but equivalent results for any other value of~$t_2$ as corollaries of the results given in this paper.
However, even \textit{stating} the results in the general case becomes much
more complicated.
For instance, it is only because
of Lemma~\ref{lem:weirdfact} that the $x$-adjoint of a wave function
just happens to be the same as its image under the bispectral
involution.
It was in order to avoid the messier formulas and the need for
additional notation that we opted to merely state the results in the
case $t_2=-\tfrac{1}{2}$.
\end{Remark}

\subsection{Almost bi-orthogonal sequences}

\begin{Definition}
For $\X=\bigl(X,Z,\vec a,\vec b\,\bigr)\in\CM_N$ let
$
\tau_{\X}(x,y)=\det\bigl(\tilde X\bigr)$ and $
q_{\X}(z)=\det\bigl(\tilde Z\bigr)
$
denote the determinants of the matrices defined in Section~\ref{sec:tilde}.
These are used to ``normalize'' the wave function as follows:
\begin{equation}
 \hat \psi_{\X}(x,y,z)=q_{\X}(z)\psi_{\X}(x,y,z)
\ \hbox{and}\
 \hat{\hat\psi}_{\X}(x,y,z)=\tau_{\X}(x,y)q_{\X}(z)\psi_{\X}(x,y,z).\label{eqn:hathat}
\end{equation}
\end{Definition}

 Note that
 multiplying by those determinants cancels a polynomial
 in the denominators so that \smash{$\hat{\hat\psi}_{\X}(x,y,z)$} is holomorphic in all
 variables and \smash{$\hat\psi_{\X}(x,y,z)$} is
 holomorphic in $z$ but may have poles in $x$.
This allows us to expand them as power series in those variables and
define sequences of quasi-polynomial functions from their
coefficients. For our purposes, it is sufficient to evaluate~\smash{$\hat\psi_{\X}$} and \smash{$\hat\psi_{\X^\flat}$} at
$y=-\tfrac{1}{2}$.

\begin{Definition}\label{def:seqs}
For $\X\in\CM_N$, define the quasi-polynomial sequences $\big\{\ff_n^{\X}(x)\big\}$ and
$\big\{\gg_n^{\X}(x)\big\}$
by
\[
\hat\psi_{\X^\flat}\bigl(x,-\tfrac{1}{2},w\bigr)=\sum_{n=0}^\infty \ff_n^{\X}(x)w^n.
\qquad\hbox{and}\qquad
\hat\psi_{\X}\bigl(x,-\tfrac{1}{2},z\bigr)=\sum_{n=0}^\infty \gg_n^{\X}(x)z^n
\]
Let the \textit{matrix of products} $\Omega_{\X}=[\omega_{mn}]$ be
 semi-infinite matrix whose entry in the $m$-th row and~$n$-th
 column is the product
 \[
\omega_{mn}=\big\langle \ff_{m-1}^{\X}(x,),\gg_{n-1}^{\X}(x)\big\rangle_{\rm H}, \qquad
m,n\geq 1.
 \]
The function $\Theta_{\X}(w,z)=\bigl\langle \psi_{\X^\flat}\bigl(x,-\tfrac{1}{2},w\bigr),\psi_{\X}\bigl(x,-\tfrac{1}{2},z\bigr)\bigr\rangle_{\rm H}$ is the
\textit{generating function for the products}
 in that
\[
\Theta_{\X}(w,z)=\sum_{m=0}^{\infty}\sum_{n=0}^{\infty}
\omega_{(m+1)(n+1)}w^mz^n.
\]
\end{Definition}

\begin{Remark}\label{rem:quasi}
By Lemma~\ref{lem:weirdfact}, the functions in the sequences $\big\{\ff_n^{\X}(x)\big\}$ and
$\big\{g_n^{\X}(x)\big\}$ are all rational functions having
$\tau_{\X}\bigl(x,-\tfrac{1}{2}\bigr)$ as their
denominator. Consequently, if one prefers to deal with sequences of
polynomials instead, you can consider the sequences
\[
\big\{\fff_n^{\X}\big\}=\bigl\{\tau_{\X}\bigl(x,-\tfrac{1}{2}\bigr)\ff_n^{\X}(x)\bigr\}\qquad\text{and}\qquad
\big\{\ggg_n^{\X}\big\}=\bigl\{\tau_{\X}\bigl(x,-\tfrac{1}{2}\bigr)\gg_n^{\X}(x)\bigr\}
\]
instead.
If we introduce the new product
\[
\langle f(x),g(x)\rangle_{\X}=\int_{-\infty}^{\infty}
f(x)g(x)\frac{{\rm e}^{-x^2/2}}{\tau_{\X}^2\bigl(x,-\tfrac{1}{2}\bigr)}{\rm d}x
\]
then $\big\langle \ff_m^{\X}(x),\gg_n^{\X}(x)\big\rangle_{\rm H}=\big\langle
\fff_m^{\X}(x),\ggg_n^{\X}(x)\big\rangle_{\X}$. Using this alternative
notation, everything said below
about the quasi-polynomial systems under the Hermite product
could equivalently be said about the polynomial sequences under that
new product. To avoid overburdening the reader with notation, we
will not write each result in both notations. (This is all somewhat
standard for exceptional orthogonal polynomials. In fact, the
function $\tau_\chi\bigl(x,-\tfrac{1}{2}\bigr)$ is essentially the same function that was
called $\eta(x)$ in~\cite{XHermite2}, with the new notation just
reflecting the fact that it can now be recognized as a KP $\tau$-function.)
\end{Remark}

\begin{Theorem}\label{thm:normgen}
The doubly-normalized wave function after evaluation at $x=w$ and $y=0$
and multiplication by $\sqrt{2\pi}$ is the generating function of the products
\begin{equation}
\Theta_{\X}(w,z)= \big\langle \hat\psistar_{\X}\bigl(x,-\tfrac{1}{2},-\tfrac{1}{2},w\bigr),\hat\psi_{\X}\bigl(x,-\tfrac{1}{2},z\bigr)\big\rangle_{\rm H}
=\sqrt{2\pi}\hat{\hat\psi}_{\X}(w,0,z).\label{eqn:ipwz}
\end{equation}
\end{Theorem}
\begin{proof}
Note first that we can pull the determinants out of the
 product since they are independent of $x$, so that the left hand
 side is equal to
 \[
\det(wI-W)\det(zI-Z)\bigl\langle
 \psistar_{\X}\bigl(x,-\tfrac{1}{2},-\tfrac{1}{2},w\bigr)\psi_{\X}\bigl(x,-\tfrac{1}{2},z\bigr)\bigr\rangle_{\rm H}.
 \]
 Then, using \eqref{eqn:scalarip} and the fact that $y=-\tfrac{1}{4}$, we know that this is equal to
 \[
\det(wI-W)\det(zI-Z) \sqrt{2\pi }
{\rm e}^{wz/2}
 \bigl(1+\vec a^{\top}\acute Z^{-1}\tilde Z^{-1}\vec b\,\bigr).
 \]

When $y=y'=-\tfrac{1}{2}$, then
 $\acute Z$ simplifies to just $\acute Z=wI-X$.
Recall $\tilde X(x)=xI-X$ where we are now explicitly writing it as a
function of the variable $x$. Then $\acute Z=\tilde X(w)$.
Now, the rational factor in the value of the product is
\begin{gather*}
\det(wI-X)\det(zI-Z)
 \bigl(1-2y'\vec a^{\top}\acute Z^{-1}\tilde Z^{-1}\vec b\,\bigr)\\
\qquad=\det(wI-X)\det(zI-Z) \bigl(1+\vec a^{\top}\bigl(2 \tilde X(w)\bigr)^{-1}\tilde Z^{-1}\vec b\,\bigr)\\
\qquad =\det(wI-X)\det(zI-Z) \bigl(1+\vec a^{\top}\tilde X^{-1}(w)\tilde Z^{-1}\vec b\,\bigr).
\end{gather*}
 Note that this is exactly the polynomial
\begin{equation*}
{\rm e}^{-wz}2^N\hat{\hat \psi}_{\X}(w,0,z)=\det(w I-X)\det(zI-Z)\bigl(1+\vec a^{\top}\tilde X^{-1}(w)\tilde
 Z^{-1}\vec b\,\bigr).\tag*{\qed}
 \end{equation*}
 \renewcommand{\qed}{}
\end{proof}

\begin{Remark}
 It is a bit confusing but also intriguing that one of the sequences of
 functions is generated by a normalized version of $\psi_{\X}(x,y,z)$
 at $y=-\tfrac{1}{2}$ while the generating function of the products is a
 normalized version of the same function at $y=0$. This
mysterious role for the KP time variables seems to suggest a deep
and not yet fully understood connection between soliton theory and the
Hermite products of the functions in these sequences.
\end{Remark}

We now associate a finite set of integers to any $\X\in\CM_{N}$. It
is found using the polynomial part of
\smash{$\hat{\hat\psi}_{\X}(x,0,z)$} and will be important in showing that the two
generated sequences of functions are almost bi-orthogonal:
\begin{Definition}\label{def:BX}
Since both polynomials have been eliminated from the denominator, the
function \smash{$\hat{\hat\psi}_{\X}$} from~\eqref{eqn:hathat} when evaluated at $y=0$ can be
written in the form
\begin{equation}
 \hat{\hat\psi}_{\X}(x,0,z)=
{\rm e}^{xz}\Bigg(\sum_{i=0}^N\sum_{j=0}^N c_{ij}x^iz^j\Bigg).\label{eqn:hathatform}
\end{equation}
 Define the set $B_{\X}\subset\mathbb{Z}$ to be the set of
 differences in the exponents for the monomials in
 \eqref{eqn:hathatform} with non-zero coefficients
$ B_{\X}=\{i-j\mid c_{ij}\not=0\}$.
\end{Definition}

For instance, suppose
\smash{$\hat{\hat\psi}_{\X}(x,0,z)={\rm e}^{xz}\bigl(1+xz^2+x^3\bigr)$}. Then
$B_{\X}=\{0,-1,3\}$. Moreover, $c_{NN}\not=0$ for every $\X\in\CM_N$
(cf.\ \eqref{eqn:psiasymptotic}), so we know that $0\in B_{\X}$
is always true.

\begin{Theorem}\label{thm:almostbiorthog}
The sequences $\big\{\ff_n^{\X}(x)\big\}$ and $\big\{\gg_n^{\X}(x)\big\}$ generated by
$\hat\psi_{\X}$ and $\hat\psi_{\X\flat}$
{\rm(}see Definition~{\rm\ref{def:seqs})} are almost bi-orthogonal in the sense that
\smash{$\big\langle \ff_m^{\X},\gg_n^{\X}\big\rangle_{\rm H}=0$} if $m-n\not\in B_{\X}$. Since~$B_{\X}$ is
a finite set, this means that the matrix of products~$\Omega_{\X}$ is a ``finite band'' matrix whose~$b$\nobreakdash-th super-diagonal is non-zero only if $b\in B_{\X}$. In
particular, the number of non-zero sub- and super-diagonals are both
bounded by the size $N$ of the Calogero--Moser matrices $\X\in\CM_N$
\end{Theorem}
\begin{proof}
Expanding the product from equation \eqref{eqn:ipwz} using bilinearity
and using \eqref{eqn:hathatform} to rewrite the right side of that
same equation, we get
\[
\sum_{m=0}^{\infty}\sum_{n=0}^{\infty}\big\langle
\ff_m^{\X}(x), \gg_n^{\X}(x)\big\rangle_{\rm H} w^mz^n
=
\left(\sum_{k=0}^{\infty}\frac{1}{k!}(wz)^k\right)\Bigg(\sum_{i=0}^N\sum_{j=0}^N c_{ij}w^iz^j\Bigg).
\]
Now suppose $m$ and $n$ are such that $\big\langle
\ff_m^{\X},\gg_n^{\X}\big\rangle_{\rm H}\not=0$. Then there must be some term from
each of the sums on the right side whose product is the monomial
$w^mz^n$. In particular, there must be some $i$, $j$, and $k$ such
that $c_{ij}\not=0$, $k+i=m$, and $k+j=n$. But then
$i-j=(i+k)-(j+k)=m-n\in B_{\X}$.
That the number of non-zero sub- and super-diagonals is bounded by $N$
follows from the fact that the polynomial in \eqref{eqn:hathatform} is
of degree at most $N$ in $x$ and $z$ and hence the elements of
$B_{\X}$ must be at least $0-N=-N$ and at most $N-0=N$.
 \end{proof}

\subsubsection{Almost bi-orthogonal example}
Consider
$
\X=\bigl(X,Z,\vec a,\vec b\,\bigr)$ where
\[
X=\left[
\begin{matrix}
 -1 & -1 \\
 1 & -\frac{3}{2} \\
\end{matrix}
\right]
,\qquad Z=\left[
\begin{matrix}
 2 & 0 \\
 0 & 3 \\
\end{matrix}
\right]
,\qquad \vec a=\left[
\begin{matrix}
 -1 \\
 -1 \\
\end{matrix}
\right]\qquad\text{and}\qquad
\vec b=\left[
\begin{matrix}
 1 \\
 1 \\
\end{matrix}
\right].
\]
Then
\begin{align*}
\hat\psi_{\X}\bigl(x,-\tfrac{1}{2},z\bigr)={}&{\rm e}^{x z-\frac{z^2}{2}}\frac{
 \bigl(2 x^2 \bigl(z^2-5
 z+6\bigr)+x \bigl(-5 z^2+21
 z-20\bigr)+5 z^2-20
 z+19\bigr)}{2 x^2-5 x+5}\\
 ={}&\sum_{n=0}^{\infty} \gg_n^{\X}(x)z^n
 =\frac{12 x^2-20 x+19}{2 x^2-5
 x+5}+\frac{2 \bigl(6 x^3-15
 x^2+20 x-10\bigr) }{2 x^2-5
 x+5}z\\
 &+\frac{\bigl(12 x^4-40
 x^3+53 x^2-30 x-9\bigr)
 }{2 \bigl(2 x^2-5
 x+5\bigr)}z^2+O\bigl(z^3\bigr)
\end{align*}
 and
\begin{align*}
 \hat \psi_{\X^\flat}\bigl(x,-\tfrac{1}{2},-1,w\bigr)
={}&
{\rm e}^{-\frac{1}{2} w (w-2 x)}\\
&\times\frac{
 \bigl(2 w^2 \bigl(2 x^2-5
 x+5\bigr)+w \bigl(10 x^2-33
 x+35\bigr)+5 \bigl(2 x^2-7
 x+10\bigr)\bigr)}{2 \bigl(2
 x^2-5 x+5\bigr)}\\
 ={}&\sum_{n=0}^{\infty} \ff_n^{\X}(x)w^n
 =\frac{5 \bigl(2 x^2-7
 x+10\bigr)}{2 \bigl(2 x^2-5
 x+5\bigr)}+\frac{ \bigl(10
 x^3-25 x^2+17 x+35\bigr)}{2
 \bigl(2 x^2-5
 x+5\bigr)}w \\
 &
 +\frac{ \bigl(10
 x^4-15 x^3-18 x^2+85
 x-30\bigr)}{4 \bigl(2 x^2-5
 x+5\bigr)}w^2+O\bigl(w^3\bigr)
\end{align*}
are the generating functions of the two sequences.

There are two interesting things about products of the form
\smash{$
\ff_m^{\X}(x)\gg_n^{\X}(x){\rm e}^{-\frac{x^2}{2}}$}.
One is that according to Theorem~\ref{thm:xint}, this product has no residues for any $m$ or $n$.

To describe the other interesting thing about such a product, we need
to find the doubly-normalized wave function
\smash{$\hat{\hat\psi}_{\X}(x,0,z)$} which is
\begin{align*}
\sqrt{2\pi} {\rm e}^{w z}
 \bigl(2 w^2 z^2-10 w^2 z+12
 w^2+5 w z^2-29 w z+40 w+5
 z^2-30 z+45\bigr).
\end{align*}
By computing $m-n$ for each monomial of the form $x^mz^n$ in the
polynomial part we find
$
B_{\X}=\{-2,-1,0,1,2\}$.
Theorem~\ref{thm:almostbiorthog} says that
\[\int_{-\infty}^{\infty}\ff_m^{\X}(x)\gg_n^{\X}(x){\rm e}^{-x^2}{\rm d}x=\big\langle \ff_m^{\X}(x),\gg_n^{\X}(x)\big\rangle_{\rm H}\not=0\]
implies $m-n$ is in that set. In this case, that means
the product will be zero whenever $|m-n|>2$.
This is reflected in the band structure of the matrix of products
\[
\Omega_{\X}=\left[
 \begin{matrix}
 45 \sqrt{\frac{\pi }{2}} &
 -15 \sqrt{2 \pi } & 5
 \sqrt{\frac{\pi }{2}} & 0 &
 0 & 0 & 0 \cdots \\
 20 \sqrt{2 \pi } & 8 \sqrt{2
 \pi } & -25 \sqrt{\frac{\pi
 }{2}} & 5 \sqrt{\frac{\pi
 }{2}} & 0 & 0 & 0 \cdots \\
 6 \sqrt{2 \pi } & 15 \sqrt{2
 \pi } & -\frac{9
 \sqrt{\frac{\pi }{2}}}{2} &
 -5 \sqrt{2 \pi } & \frac{5
 \sqrt{\frac{\pi }{2}}}{2} &
 0 & 0 \cdots \\
 0 & 6 \sqrt{2 \pi } & 5
 \sqrt{2 \pi } & -5
 \sqrt{\frac{\pi }{2}} &
 -\frac{5 \sqrt{\frac{\pi
 }{2}}}{2} & \frac{5
 \sqrt{\frac{\pi }{2}}}{6} &
 0 \cdots \\
 0 & 0 & 3 \sqrt{2 \pi } &
 \frac{5 \sqrt{\frac{\pi
 }{2}}}{3} & -\frac{47
 \sqrt{\frac{\pi }{2}}}{24}
 & -\frac{5 \sqrt{\frac{\pi
 }{2}}}{12} & \frac{5
 \sqrt{\frac{\pi }{2}}}{24}
 \cdots \\
 0 & 0 & 0 & \sqrt{2 \pi } & 0
 & -\frac{\sqrt{\frac{\pi
 }{2}}}{2} &
 -\frac{\sqrt{\frac{\pi
 }{2}}}{24}\cdots \\
 0 & 0 & 0 & 0 &
 \frac{\sqrt{\frac{\pi
 }{2}}}{2} &
 -\frac{\sqrt{\frac{\pi
 }{2}}}{12} & -\frac{23
 \sqrt{\frac{\pi }{2}}}{240}&\cdots\\
 \vdots& \vdots& \vdots& \vdots& \vdots& \vdots& \vdots&\ddots
\end{matrix}
\right].
\]
Since some of the off-diagonal entries in the matrix are non-zero and
since the two sequences are distinct, in this example the best one can
say is that the two sequences are
``almost bi-orthogonal''. That is the generic situation in
${\rm Gr}^{\rm ad}$. (Other examples below will show
cases in which one can drop the ``bi-'' and/or ``almost'' from the description.)

\subsection{Almost orthogonality}
Suppose the point of ${\rm Gr}^{\rm ad}$ corresponding to $\X=\bigl(X,Z,\vec a,\vec
b\,\bigr)\in\CM_N$
is a fixed point under the bispectral involution. (In
other words, suppose there exists an invertible $N\times N$ matrix $U$
such that $Z^{\top}=UXU^{-1}$ and $X^{\top}=UZU^{-1}$.)
Then,
$\psi_{\X}(x,y,z)=\psi_{\X^\flat}(x,y,z)$, and consequently the
generating functions of the two sequences are also equal
$
\hat\psi_{\X}\bigl(x,-\tfrac{1}{2},z\bigr)=\hat\psi_{\X^\flat}\bigl(x,-\tfrac{1}{2},z\bigr)$.
An immediate consequence of this observation is:

\begin{Theorem}\label{thm:almostorthog}
 If $\X\in\CM_N$
 has the property $\psi_{\X}(x,0,z)=\psi_{\X}(z,0,x)$,
 then $\big\{\ff_n^{\X}(x)\big\}=\big\{\gg_n^{\X}(x)\big\}$. In particular, in such a
 situation there is only one sequence and one can say that it is
 almost orthogonal to itself $($without having to mention \textit{bi}-orthogonality$)$.
\end{Theorem}

\subsubsection{Almost orthogonal example}\label{subsec:almostorthogexamp}

Consider $\X=\bigl(X,Z,\vec a,\vec b\,\bigr)$ with
\[
X=\left[\begin{matrix}
\frac{1}{2}
 \bigl(5+\sqrt{5}\bigr) & -1
 \\
 1 & \frac{1}{2}
 \bigl(5-\sqrt{5}\bigr) \end{matrix}\right],\qquad
Z=\left[\begin{matrix}
 2 & 0 \\
 0 & 3 \end{matrix}\right],\qquad
\vec a=\left[\begin{matrix}
 -1 \\
 -1 \end{matrix}\right],\qquad \hbox{and}\qquad
\vec b=\left[\begin{matrix}
1 \\
 1 \end{matrix}\right].
\]
In this case one has
\[
\psi_{\X}(x,0,z)={\rm e}^{x z} \left(1+\frac{-4 x z+10
 x+10 z+\sqrt{5}-23}{2
 \bigl(x^2-5 x+6\bigr)
 \bigl(z^2-5
 z+6\bigr)}\right)
=\psi_{\X}(z,0,x).
\]
According to Theorem~\ref{thm:almostorthog}, we then know there are
not two sequences to deal with here, but just one, generated by
\begin{gather*}
{\rm e}^{x z-\frac{z^2}{2}}\bigg(
 \frac{
2 x^2 \bigl(z^2-5 z+6\bigr)-2 x \bigl(10 z^2-48 z+55\bigr)+\bigl(49+\sqrt{5}\bigr)
 z^2
 }{2 x^2-20 x+\sqrt{5}+49}\\
 \qquad+\frac{
-5 \bigl(45+\sqrt{5}\bigr) z+7 \bigl(35+\sqrt{5}\bigr)
 }{2 x^2-20 x+\sqrt{5}+49}\bigg)
 \\
 \phantom{\qquad+}{}=\hat\psi_{\X}\bigl(x,-\tfrac{1}{2},z\bigr)
 =\sum_{n=0}^{\infty}\gg_n^{\X}(x)z^n=\sum_{n=0}^{\infty}\ff_n^{\X}(x)z^n.
\end{gather*}
 In this example, one has again that $B_{\X}=\{-2,-1,0,1,2\}$ because
\smash{$ \hat{\hat\psi}_{\X}(w,0,z)$} is
 \[
\sqrt{\frac{\pi }{2}} {\rm e}^{w z}
 \bigl(2 w^2 z^2-10 w^2 z+12
 w^2+5 w z^2-29 w z+40 w+5
 z^2-30 z+45\bigr).
 \]
 So
 the product of $\gg_m^{\X}(x)$ and $\gg_n^{\X}(x)$ should be
 zero as long as $|m-n|>2$. The quasi-polynomials in the sequence have the denominator
$ \tau_{\X}\bigl(x,-\tfrac{1}{2}\bigr)=2 x^2-20 x+\sqrt{5}+49$. A few of the numerators
in the sequence are
\begin{gather*}
\tau_{\X} \gg_1^{\X}(x)=12 x^3-120 x^2+\bigl(341+7
 \sqrt{5}\bigr) x-5
 \bigl(45+\sqrt{5}\bigr),\\
 \tau_{\X}\gg_3^{\X}(x)=2 x^5-\frac{70 x^4}{3}+\frac{1}{6}
 \bigl(509+7 \sqrt{5}\bigr)
 x^3-\frac{5}{2}
 \bigl(29+\sqrt{5}\bigr)
 x^2-\frac{1}{2} \bigl(243+5
 \sqrt{5}\bigr) x\\
 \phantom{ \tau_{\X}\gg_3^{\X}(x)=}{}+\frac{5}{2}
 \bigl(45+\sqrt{5}\bigr),
\\
\tau_{\X} \gg_4^{\X}(x)=\frac{x^6}{2}-\frac{25
 x^5}{4}+\frac{7}{24}
 \bigl(83+\sqrt{5}\bigr)
 x^4-\frac{5}{6}
 \bigl(18+\sqrt{5}\bigr)
 x^3-\frac{1}{4} \bigl(337+5
 \sqrt{5}\bigr) x^2\\
 \phantom{\tau_{\X} \gg_4^{\X}(x)=}{}+\frac{5}{4}
 \bigl(87+2 \sqrt{5}\bigr)
 x+\frac{1}{24} \bigl(147+9
 \sqrt{5}\bigr),
\end{gather*}
and
\begin{align*}
\tau_{\X} \gg_5^{\X}(x)={}&
\frac{x^7}{10}-\frac{4
 x^6}{3}+\frac{1}{120}
 \bigl(645+7 \sqrt{5}\bigr)
 x^5-\frac{5}{24}
 \bigl(5+\sqrt{5}\bigr)
 x^4-\frac{1}{12} \bigl(429+5
 \sqrt{5}\bigr) x^3
 \\
 &+\frac{5}{4}
 \bigl(41+\sqrt{5}\bigr)
 x^2+\frac{1}{8} \bigl(145+3
 \sqrt{5}\bigr) x-\frac{5}{8}
 \bigl(45+\sqrt{5}\bigr).
\end{align*}
As predicted, the Hermite products are zero when the indices differ by
at least two. For example,
 \[
 \big\langle \gg_1^{\X}(x),\gg_3^{\X}(x)\big\rangle_{\rm H}
 =6\sqrt{2\pi},
\qquad\text{but}\quad
 \big\langle \gg_1^{\X}(x),\gg_4^{\X}(x) \big\rangle_{\rm H}
= \big\langle \gg_1^{\X}(x),\gg_5^{\X}(x)\big\rangle_{\rm H}
=0.
 \]

\section{Orthogonality and the exceptional Hermites}

If \smash{$\hat{\hat\psi}_{\X}(x,0,z)=\hat{\hat\psi}_{\X}(xz,0,1)$}, then
$B_{\X}=\{0\}$ and the two generated sequences of quasi-poly\-nomials would truly be orthogonal.
We already know of some wave functions from ${\rm Gr}^{\rm ad}$ for which that
must be the case.
For the special choice
$\X=\X_{\lambda}\in\CM_N$ where $\lambda$ is a partition, the~wave
function $\hat\psi_{\X}\bigl(x,-\tfrac{1}{2},z\bigr)=z^N\psi_{\X}\bigl(x,-\tfrac{1}{2},z\bigr)$ is a generating function for the
exceptional Hermites $\big\{\hat h_n^\lambda(x)\big\}$ which are orthogonal
with respect to the Hermite inner product
\cite{KasMil,PThesis,PalusoKasman}. The next result quickly
rederives that
orthogonality as a consequence of Theorems~\ref{thm:almostbiorthog}
and~\ref{thm:almostorthog} without reference to the earlier
work or even the orthogonality of the classical Hermite polynomials.

\begin{Corollary}\label{cor:proddep}
If $\X=\X_{\lambda}\in\CM_N$ so that $\hat\psi_{\X}\bigl(x,-\tfrac{1}{2},z\bigr)$ generates the
exceptional Hermites $\big\{\hat h_n^{\lambda}(x)\big\}$, then
the three sequences $\big\{\ff_n^{\X}(x)\big\}$, $\big\{\gg_n^{\X}(x)\big\}$, and $\big\{\hat
h_n^{\lambda}(x)\big\}$ are identical.
Moreover, $B_{\X}=\{0\}$, which implies the orthogonality of the
quasi-polynomial sequence with respect to the Hermite inner product.
\end{Corollary}
\begin{proof}
 For these special points in ${\rm Gr}^{\rm ad}$, we know that there is a
 function $\gamma(\cdot)$ of one variable such that
$
 \psi_{\X}(x,0,z)=\gamma(xz)
$
 (see \cite[Section~10, Example~2]{WilsonBisp}).

This clearly implies
 $\psi_{\X}(x,0,z)=\psi_{\X}(z,0,x)$ and so by
 Theorem~\ref{thm:almostorthog}, the three sequences are identical.
 Similarly, since $\psi_{\X}(x,0,z)=\gamma(xz)$,
 the differences between
 the exponents of $w$ and $z$ in \smash{$\hat{\hat\psi}_{\X}(w,0,z)$} are
 always zero. Orthogonality follows from Theorem~\ref{thm:almostbiorthog}.
\end{proof}

Although the orthogonality of the exceptional Hermites was already known, the following
corollary of Theorem~\ref{thm:normgen} concerning the generating
function for their norms is new.
\begin{Corollary}
The doubly-normalized wave function when evaluated at $x=1$ and
$y=0$ and multiplied by $\sqrt{2\pi}$ is a generating function
for the norms of the exceptional Hermites
\[
\sqrt{2\pi}\hat{\hat\psi}_{\X_{\lambda}}(1,0,z)=\sum_{n=0}^{\infty}\big\langle \hat
h_n^{\lambda}(x),\hat h_n^{\lambda}(x)\big\rangle_{\rm H} z^n.
\]
\end{Corollary}

It would have been interesting if there were other points in ${\rm Gr}^{\rm ad}$
for which $B_{\X}=\{0\}$. However, that turns out not to be the case.

\begin{Theorem}\label{thm:onlyxh}
If $B_{\X}=\{0\}$, then there exists a partition $\lambda$ so that
$\hat\psi_{\X}(x,y,z)=\hat\psi_{\X_{\lambda}}(x,y,z)$ is a generating function
for the corresponding family of exceptional Hermites.
\end{Theorem}

 \begin{proof}
Suppose $\X=\bigl(X,Z,\vec a,\vec b\,\bigr)\in\CM_N$ has the property that
$B_{\X}=\{0\}$.
Then, by Definition~\ref{def:BX},
there is some function of one variable, $\gamma(\cdot)$, such that
\smash{$
\hat{\hat\psi}_{\X}(x,0,z)=\gamma(xz)$}.

In the standard method for producing KP wave functions from ${\rm Gr}^{\rm ad}$ by Darboux
transformation rather than using Calogero--Moser wave functions
(cf.~\cite{cmbis,WilsonBisp}, this same function can be written as
\smash{$
\hat{\hat\psi}_{\X}(x,0,z)=\tau_{\X}(x,0) K({\rm e}^{xz})
$}
where $K$ is monic ordinary differential operator in~$x$ of degree
$N$. Using the fact that $\partial_x^n({\rm e}^{xz})=z^n{\rm e}^{xz}$ one gets
immediately that the $z^N$ term is multiplied by $\tau_{\X}(x,0)$.
If $\tau_{\X}(x,0)$ has any terms of degree $m<N$, then the
doubly-normalized wave function would have a term of the form $x^mz^N$
and $m-N\not=0$ would be in $B_{\X}$. Since we~know that is not true,
$\tau_{\X}(x,0)$ must be just $x^N$. (In particular, the matrix $X$
is nilpotent.)

Since \smash{$\hat{\hat\psi}_{\X}^{\flat}(x,0,z)=\hat{\hat\psi}_{\X}(z,0,x)$},
 the same argument applies to $\X^{\flat}$ (the image under the
 bispectral involution of the CM matrices we started with). It
 follows that $Z$ also must be nilpotent.

Returning to the other standard method for producing these wave
functions (cf.~\cite{cmbis,WilsonBisp}), the fact $Z$ is nilpotent
implies that all of the finitely-supported distributions (or ``conditions'') can be chosen with support at
$z=0$. Let us suppose they are
\[
c_n(f(z))=\sum c_{nj} f^{(j)}(0)\] for~${1\leq n\leq N}$.
Then $\tau_{\X}(x,0)$ is the Wronskian of the polynomials $p_n(x)=\sum c_{nj}x^j$.
Without loss of generality, by using Gaussian elimination we
can assume that the $N$ polynomials~${p_1,\ldots,p_N}$ have distinct
highest degrees and also distinct minimal degree terms. Using
multilinearity to split the determinant into a sum of determinants of
monomials, we see that the expansion of the $\tau$-function would have
separate terms produced as the product of the highest degree terms
from each polynomial and the lowest degree terms of each polynomial.
However, we determined earlier that it is the monomial $x^N$. This
means that the highest and lowest terms in each of the polynomials is
the same.

In other words, the distributions are all of the form
$c_n(f(z))=f^{(j_n)}(0)$ for different $j_n$. But, those are exactly
the ones which correspond to matrices $\X=\X_{\lambda}$ so that the
wave function generates exceptional Hermites \cite{PThesis,PalusoKasman,WilsonCM}.
 \end{proof}

 \begin{Remark}
 Theorem~\ref{thm:onlyxh} implies that the sequences $\big\{\ff_n^{\X}(x)\big\}$ and $\big\{\gg_n^{\X}(x)\big\}$ are only
bi-orthogonal with respect to the Hermite product
$\langle\cdot,\cdot\rangle_{\rm H}$ if
\smash{$\big\{\ff_n^{\X}(x)\big\}=\big\{\gg_n^{\X}(x)\big\}=\big\{\hat h_n^\lambda(x)\big\}$} is some
sequence of exceptional Hermites.
This procedure produces no other examples of bi-orthogonality with respect to
the Hermite product. In particular, this means that the
set $\{\X\in\CM_N\mid B_{\X}=\{0\}\}$ is
exactly the same as the set $\mathcal{C}_N^{\C^\times}$, which was defined in
\cite{ScaleInvar} as the invariant set of a scaling action and shown
to be in one-to-one correspondence with certain monodromy-free
Schr\"odinger operators.
\end{Remark}

\subsection{Orthogonal example}

One way in which this example will be different than the previous two is that the matrix $X$ will depend
on a free parameter $t$ (which can be considered to be the third KP
time variable $t_3$). Let $\X=\bigl(X,Z,\vec a,\vec b\,\bigr)$ where
\[
X=\left[\begin{matrix} 0 & 0 & -3 t \\
 -1 & 0 & 0 \\
 0 & 1 & 0\end{matrix}\right],
\qquad
Z=\left[\begin{matrix} 0 & 1 & 0 \\
 0 & 0 & 1 \\
 0 & 0 & 0 \end{matrix}\right],
\qquad
\vec a= \left[\begin{matrix} 0 \\
 -3 \\
 0 \end{matrix}\right],
\qquad\text{and}\qquad
\vec b=\left[\begin{matrix} 0 \\
 1 \\
 0 \end{matrix}\right].
\]
Then the generating functions for the two sequences are
\begin{gather*}
\hat\psi_{\X}\bigl(x,-\tfrac{1}{2},z\bigr)={\rm e}^{x z-\frac{z^2}{2}}\frac{z
 \bigl(-3 t z^2+x^3 z^2-3 x^2
 z+3 x\bigr)}{x^3-3 t}=\sum_{n=0}^{\infty}\gg_n^{\X}(x,t)z^n,
\\
\hat\psi_{\X^\flat}\bigl(x,-\tfrac{1}{2},w\bigr) ={\rm e}^{xw-\frac{w^2}{2}}\frac{
 \bigl(9 t^2-3 t \bigl(w^3+3
 w+x^3+3 x\bigr)+w x \bigl(w^2
 x^2-3 w
 x+3\bigr)\bigr)}{x^3-3 t}\\[-1mm]
\hphantom{\hat\psi_{\X^\flat}\bigl(x,-\tfrac{1}{2},w\bigr)}{} =\sum_{n=0}^{\infty}\ff_n^{\X}(x,t)w^n.
\end{gather*}

It follows from Theorem~\ref{thm:xint} that
\smash{$\hat\psi_{\X^\flat}\bigl(x,-\tfrac{1}{2},w\bigr) \hat\psi_{\X}\bigl(x,-\tfrac{1}{2},z\bigr) {\rm e}^{-x^2/2}$} has no
residue at $x=c$ even where $c$ is a cube root of~$3t$, and so in
particular it is also true that \smash{$\ff_m^{\X}(x,t)\gg_n^{\X}(x,t){\rm e}^{-x^2/2}$}
has no residue for any~$m$ or~$n$ either.

To determine the set $B_{\X}$, we compute the differences of the
exponents in the polynomial part of the doubly-normalized wave
function at $t_2=0$
\[
\hat{\hat\psi}_{\X}(x,0,z)={\rm e}^{xz}\bigl(-3 t z^3+x^3 z^3-3 x^2 z^2+3 x z\bigr).
\]
Most of the monomials on the right have the same exponent on both
variables. Only the first one, therefore, would contribute a non-zero
number to $B_{\X}$. However, its coefficient would be zero if $t=0$
and only monomials with non-zero coefficient contribute to $B_{\X}$.
To go further with this example, therefore, one has to decide whether
or not $t$ is zero.

Let us first assume that $t\not=0$.
 In this case, $B_{\X}=\{-3,0\}$ and so we know that
 \[
 \big\langle
 \ff_{m}^{\X}(x,t),\gg_n^{\X}(x,t)\big\rangle_{\rm H}=\mu_n\delta_{mn}+\nu_n\delta_{m(n-3)}
 \]
 for some sequences $\{\mu_n\}$ and $\{\nu_n\}$.
 That is why the matrix of products has non-zero
 entries only on the main diagonal and the third superdiagonal{\samepage
\[
 \Omega_{\X}=\left[\begin{matrix}
0 & 0 & 0 & -3 \sqrt{2 \pi }
 t & 0 & 0 & 0 &\cdots\\
 0 & 3 \sqrt{2 \pi } & 0 & 0 &
 -3 \sqrt{2 \pi } t & 0 & 0
&\cdots \\
 0 & 0 & 0 & 0 & 0 & -3
 \sqrt{\frac{\pi }{2}} t & 0
&\cdots \\
 0 & 0 & 0 & -\sqrt{\frac{\pi
 }{2}} & 0 & 0 &
 -\sqrt{\frac{\pi }{2}} t &\cdots\\
 0 & 0 & 0 & 0 & 0 & 0 & 0 &\cdots\\
 0 & 0 & 0 & 0 & 0 &
 \frac{\sqrt{\frac{\pi
 }{2}}}{4} & 0 \\
 0 & 0 & 0 & 0 & 0 & 0 &
 \frac{\sqrt{2 \pi }}{15} &\cdots\\
\vdots& \vdots& \vdots& \vdots& \vdots& \vdots& \vdots&\ddots
 \end{matrix}\right].
\]
Since there are non-zero off diagonal entries (when
$t\not=0$), we cannot claim the sequences here are ``bi-orthogonal'',
only ``almost biorthogonal''.}

At first it might seem possible that the two sequences are identical, as in the last example, and
the prefix ``bi'' can be dropped.
However, it is clear from the fact that $\Omega_{\X}$
is not symmetric that the two sequences are different. Consequently,
in the case $t\not=0$ this example generates almost bi-orthogonal
sequences, with the terms ``almost'' or ``bi'' both being necessary to
describe this example accurately.

On the other hand, the situation changes when $t=0$. Since
$B_{\X}=\{0\}$ in that case,
 true bi-orthogonality (without the qualifier ``almost'') exists
in this case. (That is also reflected in the fact that the
off-diagonal entries in $\Omega_{\X}$ all have a factor of $t$.)

Moreover, substituting $t=0$ into the formulas above for the two
generating functions, one can see that they are equal
(apart from the change of variable $w\mapsto z$). Consequently, the
sequences they generate are also equal. There are not two sequences
here, only one which is orthogonal to itself.

When $t=0$, this example yields a sequence of quasi-polynomials which are
orthogonal with respect to the Hermite product.
Hence, as a consequence of the last theorem, they must be a family of exceptional Hermite polynomials.
In fact, one can check that in the case $t=0$ these are the Calogero--Moser matrices
corresponding to the partition $\lambda=(\lambda_1,\lambda_2,\lambda_3)=(2,1,0)$ \cite{PThesis,PalusoKasman,WilsonCM}.

\section{Spin Calogero--Moser and matrix orthogonality}

So far, we have considered Calogero--Moser matrices $X$ and $Z$ satisfying
$\operatorname{rank}([X,Z]-I)=1$. If their commutator differs from the
identity by a matrix of rank $r>1$, then the same formula for the wave
function gives a solution to the non-commutative KP hierarchy whose
poles move according to a \textit{spin generalized} Calogero--Moser
system \cite{BGK}. It turns out that all of the results above also
generalize to that case so as to produce orthogonal $r\times r$ matrix
functions, with the proofs all being essentially the
same. Here we briefly state the main results and provide an example
to illustrate.

\begin{Definition}\label{def:spin}
Suppose $\X=(X,Z,A,B)$ consists of $N\times N$ matrices $X$ and $Z$
and $N\times r$ matrices $A$ and $B$ satisfying the equation
$[X,Z]-I_N=BA^{\top}$. Define the associated normalized wave
functions
\[
\hat\psi_{\X}(x,y,z)={\rm e}^{xz+yz^2}\bigl(I_r+A^\top(xI_N+2yZ-X)^{-1}(zI_N-Z)^{-1}B\bigr)\det(zI_N-Z)
 \]
 and
\smash{$
\hat{\hat\psi}_{\X}(x,y,z)=\hat\psi_{\X}(x,y,z)\det(xI_N+2yZ-X)$}.
The function \smash{$\hat{\hat\psi}_{\X}(x,0,z)$} can still be written in form
\eqref{eqn:hathatform} with constant non-zero $r\times r$ matrices~$c_{ij}$ and $B_{\X}$ is still defined to be the set of values $i-j$
that appear.
\end{Definition}
 For any $\X$ as in Definition~\ref{def:spin},
 $\X^{\flat}=\bigl(Z^\top,X^\top,B,A\bigr)$ is another 4-tuple of matrices
 satisfying the same rank $r$ condition. An appropriate product of their
 associated normalized wave functions at $y=-\tfrac{1}{2}$ with the weight
 function ${\rm e}^{-x^2/2}$ again has nice properties.

\begin{Theorem}\label{thm:spin}
The product
 \smash{$
P(x,y,w,z)= \psi_{\X^\flat}^\top\bigl(x,-\tfrac{1}{2},w\bigr)\psi_{\X}\bigl(x,-\tfrac{1}{2},z\bigr){\rm e}^{-x^2/2}
$} satisfies
\[
\oint P(x,y,w,z){\rm d}x=0
\qquad\hbox{and}\qquad
\int_{-\infty}^{\infty} P(x,y,w,z){\rm d}x=\sqrt{2\pi}\psi_{\X}(w,0,z).
\]
Consequently, if we define the sequences of matrix quasi-polynomials
$\big\{\ff_n^{\X}(x)\big\}$ and $\big\{\gg_n^{\X}(x)\big\}$ as the coefficients in the
power series expansions of \smash{$\hat\psi_{\X^\flat}\bigl(x,-\tfrac{1}{2},w\bigr)$} and
\smash{$\hat\psi_{\X}\bigl(x,-\tfrac{1}{2},z\bigr)$}, respectively, then
\begin{equation}\label{eqn:matprod}
\int_{-\infty}^{\infty} (\ff_m^{\X}(x))^\top
\gg_n^{\X}(x){\rm e}^{-\frac{x^2}2}{\rm d}x=0
\end{equation}
if $m-n\not\in B_{\X}$. The sequences are therefore always ``almost
bi-orthogonal'', are almost orthogonal if
\smash{$\hat\psi_{\X}(z,0,x)=\hat\psi_{\X^\flat}^{\top}(x,0,z)$}, and are bi-orthogonal
if \smash{$\hat{\hat\psi}_{\X}(x,0,z)=\hat{\hat\psi}_{\X}(xz,0,1)$}.
\end{Theorem}

\subsection{Matrix orthogonality example}
The case
\[
X=\left[\begin{matrix}
 0 & 0 & 0 & 0 \\
 0 & 0 & 0 & 0 \\
 1 & 0 & 1 & 1 \\
 0 & 1 & 1 & 1
 \end{matrix}\right],\qquad
Z=\left[\begin{matrix}
 0 & 0 & 1 & 0 \\
 0 & 0 & 0 & 1 \\
 0 & 0 & 0 & 0 \\
 0 & 0 & 0 & 0 \end{matrix}\right],\qquad
A=\left[\begin{matrix}
 -2 & 0 \\
 0 & -2 \\
 -1 & -1 \\
 -1 & -1 \end{matrix}\right],\qquad
B=\left[\begin{matrix}
 1 & 0 \\
 0 & 1 \\
 0 & 0 \\
 0 & 0 \end{matrix}\right]
\]
is interesting because then
\[
\hat{\hat\psi}_{\X}(x,0,z)={\rm e}^{xz}\left[\begin{matrix} x^4 z^4-2 x^3 z^4-2 x^3 z^3+3 x^2 z^3 & -x^2 z^3 \\
 -x^2 z^3 & x^4 z^4-2 x^3 z^4-2 x^3 z^3+3 x^2 z^3\end{matrix}\right].
\]
Note that the powers of $x$ are always equal to or one less than the
powers of $z$, which means $B_{\X}=\{-1,0\}$. Therefore, the
integral \eqref{eqn:matprod} will be zero whenever $m>n$ or when
$m<n-1$. That is true, for instance, for
\begin{gather*}
\ff_5^{\X}(x)=\!\left[\begin{matrix}
\dfrac{-3 x^6\hspace{-0.75pt}+\hspace{-0.75pt}6 x^5\hspace{-0.75pt}+\hspace{-0.75pt}3 x^4 \hspace{-0.75pt}-\hspace{-0.75pt}4 x^3\hspace{-0.75pt}-\hspace{-0.75pt}9 x^2\hspace{-0.75pt}+\hspace{-0.75pt}6 x\hspace{-0.75pt}+\hspace{-0.75pt}9}{6 \bigl(x^4\hspace{-0.75pt}-\hspace{-0.75pt}2 x^3\hspace{-0.75pt}-\hspace{-0.75pt}2 x^2\hspace{-0.75pt}+\hspace{-0.75pt}2
 x\hspace{-0.75pt}+\hspace{-0.75pt}1\bigr)}\!\hspace{-0.5pt} & \dfrac{-x^6\hspace{-0.75pt}+\hspace{-0.75pt}2 x^5\hspace{-0.75pt}+\hspace{-0.75pt}x^4\hspace{-0.75pt}-\hspace{-0.75pt}4 x^3\hspace{-0.75pt}-\hspace{-0.75pt}3 x^2\hspace{-0.75pt}-\hspace{-0.75pt}6 x\hspace{-0.75pt}+\hspace{-0.75pt}3}{6 \bigl(x^4\hspace{-0.75pt}-\hspace{-0.75pt}2
 x^3\hspace{-0.75pt}-\hspace{-0.75pt}2 x^2\hspace{-0.75pt}+\hspace{-0.75pt}2 x\hspace{-0.75pt}+\hspace{-0.75pt}1\bigr)} \vspace{1mm}\\
 \dfrac{-x^6\hspace{-0.75pt}+\hspace{-0.75pt}2 x^5\hspace{-0.75pt}+\hspace{-0.75pt}x^4\hspace{-0.75pt}-\hspace{-0.75pt}4 x^3\hspace{-0.75pt}-\hspace{-0.75pt}3 x^2\hspace{-0.75pt}-\hspace{-0.75pt}6 x\hspace{-0.75pt}+\hspace{-0.75pt}3}{6 \bigl(x^4\hspace{-0.75pt}-\hspace{-0.75pt}2 x^3\hspace{-0.75pt}-\hspace{-0.75pt}2 x^2\hspace{-0.75pt}+\hspace{-0.75pt}2
 x\hspace{-0.75pt}+\hspace{-0.75pt}1\bigr)} & \dfrac{-3 x^6\hspace{-0.75pt}+\hspace{-0.75pt}6 x^5\hspace{-0.75pt}+\hspace{-0.75pt}3 x^4\hspace{-0.75pt}-\hspace{-0.75pt}4 x^3\hspace{-0.75pt}-\hspace{-0.75pt}9 x^2\hspace{-0.75pt}+\hspace{-0.75pt}6 x\hspace{-0.75pt}+\hspace{-0.75pt}9}{6
 \bigl(x^4\hspace{-0.75pt}-\hspace{-0.75pt}2 x^3\hspace{-0.75pt}-\hspace{-0.75pt}2 x^2\hspace{-0.75pt}+\hspace{-0.75pt}2 x\hspace{-0.75pt}+\hspace{-0.75pt}1\bigr)}
 \end{matrix}\right]
\end{gather*}
and
\[
\gg_4^{\X}(x)=\left[\begin{matrix}
 \dfrac{-x^4+x^3+x+1}{x^4-2 x^3-2 x^2+2 x+1} & -\dfrac{x^3+x}{x^4-2
 x^3-2 x^2+2 x+1} \vspace{1mm}\\
 -\dfrac{x^3+x}{x^4-2 x^3-2 x^2+2 x+1} & \dfrac{-x^4+x^3+x+1}{x^4-2
 x^3-2 x^2+2 x+1}\end{matrix}\right].
\]
It is not immediately clear that the integral in \eqref{eqn:matprod}
would be well-defined, let alone that it would be zero, for these
functions. However, the results above guarantee that is.

\section{Concluding remarks}

Beginning with the KP hierarchy written in its integral form
\eqref{eqn:KPasintegral}, we showed that each coefficient in the power
series expansion of any normalized ${\rm Gr}^{\rm ad}$ wave function
$\hat\psi_{\X}\bigl(x,-\tfrac{1}{2},z\bigr)$ is orthogonal with respect to the Hermite
product to all but a finite number of coefficients from the expansion
of its bispectral dual $\hat\psi_{\X^\flat}\bigl(x,-\tfrac{1}{2},z\bigr)$ (Theorem~\ref{thm:almostbiorthog}). The same wave
function evaluated at $t_2=0$ encodes the
pairwise products of elements from those two sequences and, in particular, reveals which of them can be
non-zero (Theorem~\ref{thm:normgen}). The orthogonality of
the exceptional Hermites is recovered as a special case of this
construction (Corollary~\ref{cor:proddep}), thereby providing
a context for understanding why generating
functions for exceptional Hermites are KP wave functions \cite{KasMil}.

The important roles played by the KP flows, the bispectral involution,
and the bilinear equations of the KP hierarchy written in integral
form in deriving these results suggest it is not merely a coincidence
that some KP wave functions are generating functions for exceptional
orthogonal polynomials. An interesting question that this raises is
whether similar results could be obtained by considering $t_n\not=0$
for $n\geq 3$ or other KP solutions outside of ${\rm Gr}^{\rm ad}$.

Another question raised (but not addressed) by this paper is what
linear operators have interesting actions on the sequences from Definition~\ref{def:seqs}.
Each family of exceptional Hermites is known to be an eigenfunction
for a second-order differential operator in $x$ having an eigenvalue
depending on the index \textit{and} an eigenfunction for a difference
operator having an eigenvalue which is a polynomial in $x$ \cite{XHermite2}. It would
be interesting to know whether the sequences of quasi-polynomials
$\big\{\ff_n^{\X}(x)\big\}$ and $\big\{\gg_n^{\X}(x)\big\}$ \textit{other} than
exceptional Hermites are also eigenfunctions of any linear operators.

\subsection*{Acknowledgements} The authors thank Dan
Maroncelli and Nick Davidson at the College of Charleston, Maxim
Derevyagin at the University of Connecticut, and the anonymous referees for helpful conversations and advice.

\pdfbookmark[1]{References}{ref}
\LastPageEnding

\end{document}